\documentclass[twocolumn]{aastex631}
\usepackage{graphicx,multirow,hyperref,url,color,xspace}
\usepackage{soul}
\usepackage{amsmath,amssymb}

\shorttitle{}
\shortauthors{L.H.X et al.}

\usepackage{booktabs}
\begin{document}

\title{Transitions and Origin of the Type-B Quasi-Periodic Oscillation in the Black Hole X-ray Binary MAXI~ J1348--630}
\author{H.X.Liu*}
\affiliation{Key Laboratory for Particle Astrophysics, Institute of High Energy Physics, Chinese Academy of Sciences, 19B Yuquan Road, Beijing 100049, People's Republic of China}
\affiliation{University of Chinese Academy of Sciences, Chinese Academy of Sciences, Beijing 100049, People's Republic of China}
\email{liuhexin@ihep.ac.cn}

\correspondingauthor{Y. Huang}
\email{huangyue@ihep.ac.cn}
\correspondingauthor{Q.C.Bu}
\email{bu@astro.uni-tuebingen.de}

\author{Y. Huang*}
\affiliation{Key Laboratory for Particle Astrophysics, Institute of High Energy Physics, Chinese Academy of Sciences, 19B Yuquan Road, Beijing
100049, People's Republic of China}
\affiliation{University of Chinese Academy of Sciences, Chinese Academy of Sciences, Beijing 100049, People's Republic of China}

\author{Q.C.Bu*}
\affiliation{Institut f\"ur Astronomie und Astrophysik, Kepler Center for Astro and Particle Physics, Eberhard Karls Universit\"at, Sand 1, 72076 T\"ubingen, Germany}

\author{W. Yu}
\affiliation{Key Laboratory for Particle Astrophysics, Institute of High Energy Physics, Chinese Academy of Sciences, 19B Yuquan Road, Beijing
100049, People's Republic of China}
\affiliation{University of Chinese Academy of Sciences, Chinese Academy of Sciences, Beijing 100049, People's Republic of China}
\author{Z.X.Yang}
\affiliation{Key Laboratory for Particle Astrophysics, Institute of High Energy Physics, Chinese Academy of Sciences, 19B Yuquan Road, Beijing
100049, People's Republic of China}
\affiliation{University of Chinese Academy of Sciences, Chinese Academy of Sciences, Beijing 100049, People's Republic of China}

\author{L. Zhang}
\affiliation{Key Laboratory for Particle Astrophysics, Institute of High Energy Physics, Chinese Academy of Sciences, 19B Yuquan Road, Beijing
100049, People's Republic of China}

\author{L.D.Kong}
\affiliation{Key Laboratory for Particle Astrophysics, Institute of High Energy Physics, Chinese Academy of Sciences, 19B Yuquan Road, Beijing
100049, People's Republic of China}
\affiliation{University of Chinese Academy of Sciences, Chinese Academy of Sciences, Beijing 100049, People's Republic of China}
\author{G.C.XIAO}
\affiliation{Purple Mountain Observatory, Chinese Academy of Sciences,Nanjing 210034, People's Republic of China}
\author{J.L.Qu}
\affiliation{Key Laboratory for Particle Astrophysics, Institute of High Energy Physics, Chinese Academy of Sciences, 19B Yuquan Road, Beijing
100049, People's Republic of China}
\affiliation{University of Chinese Academy of Sciences, Chinese Academy of Sciences, Beijing 100049, People's Republic of China}

\author{S.N.Zhang}
\affiliation{Key Laboratory for Particle Astrophysics, Institute of High Energy Physics, Chinese Academy of Sciences, 19B Yuquan Road, Beijing 100049, People's Republic of China}
\affiliation{University of Chinese Academy of Sciences, Chinese Academy of Sciences, Beijing 100049, People's Republic of China}

\author{S.Zhang}
\affiliation{Key Laboratory for Particle Astrophysics, Institute of High Energy Physics, Chinese Academy of Sciences, 19B Yuquan Road, Beijing 100049, People's Republic of China}
\affiliation{University of Chinese Academy of Sciences, Chinese Academy of Sciences, Beijing 100049, People's Republic of China}

\author{L.M.Song}
\affiliation{Key Laboratory for Particle Astrophysics, Institute of High Energy Physics, Chinese Academy of Sciences, 19B Yuquan Road, Beijing 100049, People's Republic of China}
\affiliation{University of Chinese Academy of Sciences, Chinese Academy of Sciences, Beijing 100049, People's Republic of China}

\author{S.M.JIA}
\affiliation{Key Laboratory for Particle Astrophysics, Institute of High Energy Physics, Chinese Academy of Sciences, 19B Yuquan Road, Beijing 100049, People's Republic of China}
\affiliation{University of Chinese Academy of Sciences, Chinese Academy of Sciences, Beijing 100049, People's Republic of China}

\author{X.MA}
\affiliation{Key Laboratory for Particle Astrophysics, Institute of High Energy Physics, Chinese Academy of Sciences, 19B Yuquan Road, Beijing 100049, People's Republic of China}
\affiliation{University of Chinese Academy of Sciences, Chinese Academy of Sciences, Beijing 100049, People's Republic of China}

\author{L.TAO}
\affiliation{Key Laboratory for Particle Astrophysics, Institute of High Energy Physics, Chinese Academy of Sciences, 19B Yuquan Road, Beijing 100049, People's Republic of China}
\affiliation{University of Chinese Academy of Sciences, Chinese Academy of Sciences, Beijing 100049, People's Republic of China}

\author{M.Y.GE}
\affiliation{Key Laboratory for Particle Astrophysics, Institute of High Energy Physics, Chinese Academy of Sciences, 19B Yuquan Road, Beijing 100049, People's Republic of China}
\affiliation{University of Chinese Academy of Sciences, Chinese Academy of Sciences, Beijing 100049, People's Republic of China}

\author{Q.Z.LIU}
\affiliation{Purple Mountain Observatory, Chinese Academy of Sciences,Nanjing 210034, People's Republic of China}

\author{J.Z.YAN}
\affiliation{Purple Mountain Observatory, Chinese Academy of Sciences,Nanjing 210034, People's Republic of China}

\author{R.C.MA}
\affiliation{Key Laboratory for Particle Astrophysics, Institute of High Energy Physics, Chinese Academy of Sciences, 19B Yuquan Road, Beijing 100049, People's Republic of China}
\affiliation{University of Chinese Academy of Sciences, Chinese Academy of Sciences, Beijing 100049, People's Republic of China}

\author{X.Q.REN}
\affiliation{Key Laboratory for Particle Astrophysics, Institute of High Energy Physics, Chinese Academy of Sciences, 19B Yuquan Road, Beijing 100049, People's Republic of China}
\affiliation{University of Chinese Academy of Sciences, Chinese Academy of Sciences, Beijing 100049, People's Republic of China}
\author{D.K.ZHOU}
\affiliation{Key Laboratory for Particle Astrophysics, Institute of High Energy Physics, Chinese Academy of Sciences, 19B Yuquan Road, Beijing 100049, People's Republic of China}
\affiliation{University of Chinese Academy of Sciences, Chinese Academy of Sciences, Beijing 100049, People's Republic of China}

\author{T.M.LI}
\affiliation{Key Laboratory for Particle Astrophysics, Institute of High Energy Physics, Chinese Academy of Sciences, 19B Yuquan Road, Beijing 100049, People's Republic of China}
\affiliation{University of Chinese Academy of Sciences, Chinese Academy of Sciences, Beijing 100049, People's Republic of China}

\author{B.Y.WU}
\affiliation{Key Laboratory for Particle Astrophysics, Institute of High Energy Physics, Chinese Academy of Sciences, 19B Yuquan Road, Beijing 100049, People's Republic of China}
\affiliation{University of Chinese Academy of Sciences, Chinese Academy of Sciences, Beijing 100049, People's Republic of China}

\author{Y.C.XU}
\affiliation{Key Laboratory for Particle Astrophysics, Institute of High Energy Physics, Chinese Academy of Sciences, 19B Yuquan Road, Beijing 100049, People's Republic of China}
\affiliation{University of Chinese Academy of Sciences, Chinese Academy of Sciences, Beijing 100049, People's Republic of China}

\author{Y.F.DU}
\affiliation{Key Laboratory for Particle Astrophysics, Institute of High Energy Physics, Chinese Academy of Sciences, 19B Yuquan Road, Beijing 100049, People's Republic of China}
\affiliation{University of Chinese Academy of Sciences, Chinese Academy of Sciences, Beijing 100049, People's Republic of China}

\author{Y.C.FU}
\affiliation{Key Laboratory for Particle Astrophysics, Institute of High Energy Physics, Chinese Academy of Sciences, 19B Yuquan Road, Beijing 100049, People's Republic of China}
\affiliation{University of Chinese Academy of Sciences, Chinese Academy of Sciences, Beijing 100049, People's Republic of China}

\author{Y.X.XIAO}
\affiliation{Key Laboratory for Particle Astrophysics, Institute of High Energy Physics, Chinese Academy of Sciences, 19B Yuquan Road, Beijing 100049, People's Republic of China}
\affiliation{University of Chinese Academy of Sciences, Chinese Academy of Sciences, Beijing 100049, People's Republic of China}

\author{G.Q. Ding}
\affiliation{Xinjiang Astronomical Observatory, Chinese Academy of Sciences,150, Science 1-Street, Urumqi, Xinjiang 830011, China}

\author{X.X. YU}
\affiliation{Key Laboratory for Particle Astrophysics, Institute of High Energy Physics, Chinese Academy of Sciences, 19B Yuquan Road, Beijing 100049, People's Republic of China}
\affiliation{University of Chinese Academy of Sciences, Chinese Academy of Sciences, Beijing 100049, People's Republic of China}
\begin{abstract}
The fast transitions between different types of quasi-periodic oscillations (QPOs) are generally observed in black hole transient sources (BHTs). We present a detailed study on the timing and spectral properties of the transitions of type-B QPOs in MAXI~J1348--630, observed by \emph{Insight}-HXMT. The fractional rms variability--energy relationship and energy spectra reveal that type-B QPOs probably originate from jet precession. Compared to weak power-law dominated power spectrum, when type-B QPO is present, the corresponding energy spectrum shows an increase in Comptonization component and the need for {\tt\string xillverCp} component, and a slight increase of height of the corona when using {\tt\string relxilllp} model. Therefore, we suggest that a coupled inner disk-jet region is responsible for the observed type-B QPOs transitions. The time scale for the appearance/disappearance of type-B QPOs is either long or short (seconds), which may indicate an instability of disk-jet structure. For these phenomena, we give the hypothesis that the Bardeen-Petterson effect causes disk-jet structure to align with BH spin axis, or that the disappearance of small-scale jets bound by the magnetic flux tubes lead to the disappearance of type-B QPOs.
We observed three events regarding the B/C transitions, one of which occurred in a short time from $\sim 9.2$ Hz (C) to $\sim 4.8$ Hz (B).
The energy spectral analysis for the other two transitions shows that when type-C QPO is present, the Comptonization flux is higher, the spectrum is harder and the inner radius of disk changes insignificantly. We suggest that type-C QPOs probably originate from relatively stronger jets or corona.

\end{abstract}

\section{Introduction} \label{sec:intro}
Most of the black hole X-ray binaries (BHXBs) in the Galaxy are discovered as transients whose spectral and timing properties change with time. During a typical outburst, a BHXB goes through a transition from the low/hard (LHS) to the high/soft (HSS) state through relatively short--lived intermediate states, which are known as hard intermediate state (HIMS) and soft intermediate state (SIMS) \citep{2010LNP...794...53B,2006ARA&A..44...49R,2009MNRAS.400.1603M}. The rapid flux variations found in the X-ray light curves of BHXBs are generally considered to be a feature of the inner accretion flow in the vicinity of black holes. Low frequency Quasi Periodic Oscillations (LFQPOs), characterized by discrete peaks in the Power Density Spectra (PDS), are commonly observed in the 0.01--30 Hz range in BHXBs. Based on the properties of PDS in several typical BHXBs, XTE J1550--564, GX 339--4, and XTE~J1859+226 \citep{1999ApJ...526L..33W,2001ApJS..132..377H,2011MNRAS.418.2292M,2005AIPC..797..225C}, LFQPOs are classified into three types, namely type-A, B, and C. Type-A QPOs are rarely detected during the HSS, and they are characterized by a weak and broad peak in the PDS. During the transition  state---SIMS, type-B QPOs are usually observed, and typically have a rather high amplitude in the PDS of a power law shape with a centroid frequency around 4--6 Hz. Generally seen in the LHS and HIMS, type-C QPOs are characterized as a strong narrow peak with broadband noise in the PDS, which may be accompanied by harmonics and possible sub-harmonics \citep{2004A&A...426..587C}. 

Through timing analysis, rapid transitions between two types of QPOs (type-B QPOs transit to other types or disappear), have been observed in a few BHXBs. As reported by \citet{1991ApJ...383..784M}, \citet{2003A&A...412..235N} and \citet{2011MNRAS.418.2292M}, the transitions of type-B QPOs are often observed in GX~339--4. In XTE~J1550--564 \citep{2001ApJS..132..377H,2016ApJ...823...67S}, transition events related to type-B/A QPOs variations were observed along with the change of energy spectra. Similar rapid transitions involving type-B QPOs and spectral variations in XTE~J1859+226 were reported by \citet{2004A&A...426..587C} and \citet{2013ApJ...775...28S}. In XTE~J1817--330, \citet{2012A&A...541A...6S} found that a type-B QPO $\sim$ 6 Hz QPO switches to a type-A QPO with a decrease in flux. Among all the outbursts of H1743-322 observed by Rossi X-ray Timing Explorer (\emph{RXTE}), not only were type B/A QPOs transitions observed, \citep{2005ApJ...623..383H} but also three B/C transitions \citep{2021ApJ...911..127S}. For MAXI~J1535--571, \citet{2018ApJ...866..122H} found type-B QPOs $\sim$ 10 Hz in correspondence with flux changes in X--ray, and \citet{2018ApJ...865L..15S} reported a weak type A/B QPO transition with no sudden flux change. The type C/B QPO transitions were also found in GRS 1915+105 \citep{2008MNRAS.383.1089S} and MAXI J1820+070 \citep{2020ApJ...891L..29H}.

Although LFQPOs have been discovered for decades, there is still no consensus for their physical explanations. Some promising models have been proposed based on observational data, which can be divided into two categories: the geometric effects of the hot inner/outer flows and the instability of accretion flows. The relativistic precession model was firstly proposed by \citet{1998ApJ...492L..59S} to explain the origin and the behaviour of the LFQPO and kHz QPOs in neutron star X-ray binaries.
This model has been continuously modified \citep{2009MNRAS.397L.101I,2015MNRAS.448.1298P,2017MNRAS.464.2643V,2021NatAs...5...94M} later and applied to BHXBs. In these improved models, the QPOs are from Lense-Thirring  (L-T) precession of a hot inner flow (a radially extended corona or jet). These observational evidences support the geometric explanations for LFQPOs \citep{2006ApJ...642..420S,2009MNRAS.397L.101I,2016MNRAS.461.1967I}.
In the other kind of model, the QPOs come from intrinsic instabilities in the accretion flow, for example, magneto-acoustic wave propagation \citep{2004ApJ...612..988T,2010MNRAS.404..738C}, or the accretion-ejection instability (AEI)---a product of instability accretion model \citep{1999A&A...349.1003T}.

It is thought that the different types of QPOs observed in the BHXBs systems are associated with different physical mechanisms in the inner region of the accretion disk. However, the inner disk variation could not be conclusively demonstrated previously, which may be due to the energy band of \emph{RXTE}. Therefore, the transition mechanism between different types of QPOs has not been clear.  Using the data of \emph{Insight}-HXMT, with a wider energy band (1 keV--250 keV), it is now possible to obtain more information and to better understand the changes that occur in the inner disk regions as well as in the high-energy radiation regions where the LFQPOs and their transitions are observed.

MAXI J1348--630 was discovered as a X-ray binary by the X-ray telescope \textit{MAXI}/GSC on January 26, 2019 \citep{2019ATel12425....1Y}.  \citet{2020ApJ...897....3J} estimated the mass of the BH to be $9.1^{+1.6}_{-1.2}$ $M_{\sun}$. They also found that the viscous timescale in this outburst is $\sim$3.5 days. The distance of the source was estimated as $\sim$2.2 kpc by \citet{2020MNRAS.tmpL.236C}. \citet{2020ApJ...899L..20T} proposed that the black hole is more massive if the disk is inclined and the black hole is spinning. These results suggest that MAXI~J1348--630 may host a relatively massive black hole among the known BHXBs in our Galaxy. \citet{2019ATel12497....1C} reported that a radio source is detected at a position consistent with the location of MAXI~J1348--630, displaying, on 2019-02-09, a preliminary flux density of 520.3 $\pm$ 5.0 mJy.

The QPOs transitions between different typical QPOs were also observed in the 2019 outburst of MAXI~J1348--630. For this outburst, \citet{2021MNRAS.505.3823Z} have analyzed four fast appearance/disappearance of type-B QPO using the data from \emph{NICER}, in the energy band of 0.5--10 keV.

\emph{Insight}-HXMT has observed several transitions involving the type-B QPOs with its three main detectors: LE, ME, and HE, covering a broad-band energy (1--250 keV). Using the data from \emph{Insight}-HXMT, the detailed timing and spectral analyses for the transitions given in this work may provide more information to understand the underlying physical processes.
The paper is organized as follows: the data reduction is introduced in Section \ref{data}, and we present the timing and spectral results in Section \ref{Results}; in Section \ref{discussion}, we present the summary and discussions; finally, we present conclusions in Section \ref{conclusion}.

\section{Observations and Data Reduction}
\label{data}
\emph{Insight}-HXMT, China's first X-ray astronomical satellite which was successfully launched on June 15, 2017, carries three sets of main instruments (LE/ME/HE, short for the Low/Medium/High Energy X-ray Telescope respectively). More details about \emph{Insight}-HXMT can be found in \citet{2020SCPMA..63x9502Z}, \citet{2020SCPMA..63x9505C} (LE), \citet{2020SCPMA..63x9504C} (ME) and \citet{2020SCPMA..63x9503L} (HE). MAXI J1348--630 was observed regularly with \emph{Insight}-HXMT from January 27 2019 to May 15. We analyzed the observations in the SIMS around the peak of this outburst (see the Figure \ref{fig:evolution}) in this work.

The \emph{Insight}-HXMT Data Analysis software (HXMTDAS, v2.04) is used to process and filter the data. Since the background and some particle events do not have significant influences on the PDS, the screening criteria of good time-intervals (GTIs) in timing analysis can be loosened. Therefore, the light curves to produce dynamic power spectra (DPS) are generated with the following criteria: the offset for the point position $ < 0.06^\circ$, the elevation angle (ELV) $ > 10^\circ$. The selection of GTIs is more demanding when extracting the energy spectra: ELV $>10^\circ$; the geometric cutoff rigidity (COR)  $>6^\circ$; the offset for the point position $<0.04^\circ$; data are used at least 300 s before and after the South Atlantic Anomaly (SAA) passage. The backgrounds are estimated with the official tools: LEBKGMAP \citep{2020JHEAp..27...24L}, MEBKGMAP \citep{2020JHEAp..27...44G} (ME)and HEBKGMAP \citet{2020JHEAp..27...14L} in version 2.0.6.

The energy bands, adopted for energy spectral analysis are 1--1.8, 2--10 keV (LE), 10--35 keV (ME) and 35--100 keV (HE) in this work. The XSPEC software package v12.11 \citep{1996ASPC..101...17A} is used to fit the spectra. Uncertainty estimated for each spectral parameter is for 90\% confidence level, and a systematic error of 1\% is added. Detailed discussions of its calibrations are given in \citet{2020JHEAp..27...64L}.

\section{Data Analysis and Results}
\label{Results}
\subsection{Timing Analysis and Results}
\label{Timing Results}
\begin{figure}[!htbp]
    \centering
    \includegraphics[scale=0.5]{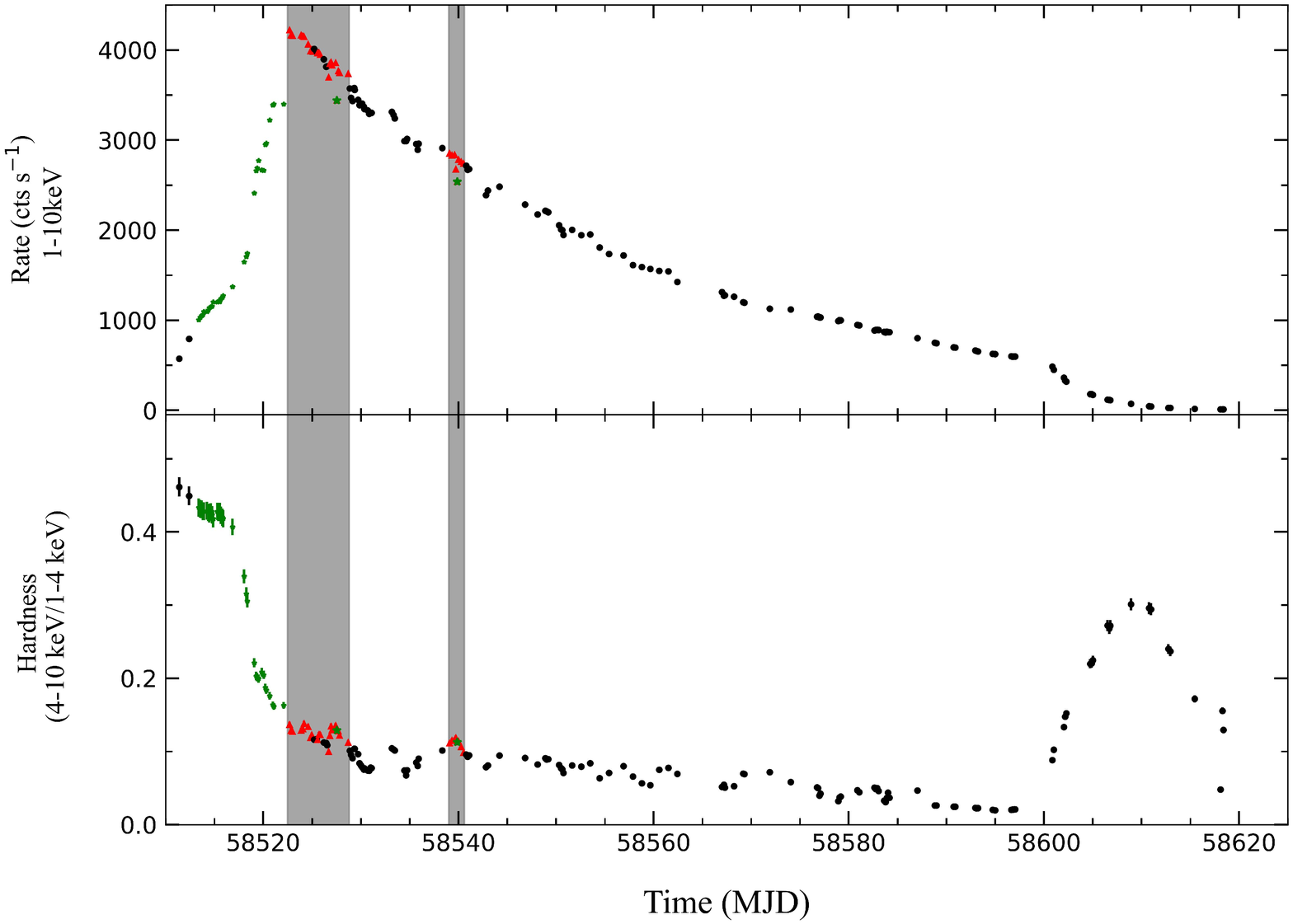}
    \caption{\emph{Insight}-HXMT/LE (1--10 keV) light curve (top panel) and the hardness (bottom panel) for MAXI~J1348--630 during its 2019 outburst. Each point corresponds to one exposure. Hardness is defined as the ratio of LE count rate in 4--10 keV and 1--4 keV. Green points correspond to observations with a type-C QPO in the PDS, red points correspond the observations with a type-B QPO and black points mark all the other observations that do not show low-frequency QPOs.}
    \label{fig:evolution}
\end{figure}

Using the archived data of \emph{Insight}-HXMT observations of MAXI~J1348--630, we plot the evolution of its light curves and hardness in Figure \ref{fig:evolution}, with each point in the diagram representing an observation.
The top panel represents the count rate of LE (1--10 keV), while the bottom panel shows the hardness ratio that is defined as count rates ratio between 4--10 keV and 2--4 keV. According to the the spectral analysis, \citet{2022arXiv220111919Z} have defined four states for this outburst. Huang et al. (2022) (In prepation) studied in detail the timing properties in the 2019 outburst. They found that the observational properties are consistent with BHXBs and the type-C QPOs frequency varied between 0.26 and 7.31 Hz with decreasing hardness. 

The appearance of the type-B QPO indicates the source was in its SIMS, as shown in the shaded area of Figure \ref{fig:evolution}.
In this work, we focus on studying the transitions of QPOs during the SIMS of this outburst, while the detailed information are listed in Table \ref{QPO}. For our timing analysis, we compute the power density spectra (PDS) with a time resolution of 1/128 s using 16 s data segments (corresponding to a Nyquist frequency of 64 Hz) for the observation listed in Table \ref{QPO}. 

\begin{figure*}
    \centering
    \includegraphics[scale=0.5]{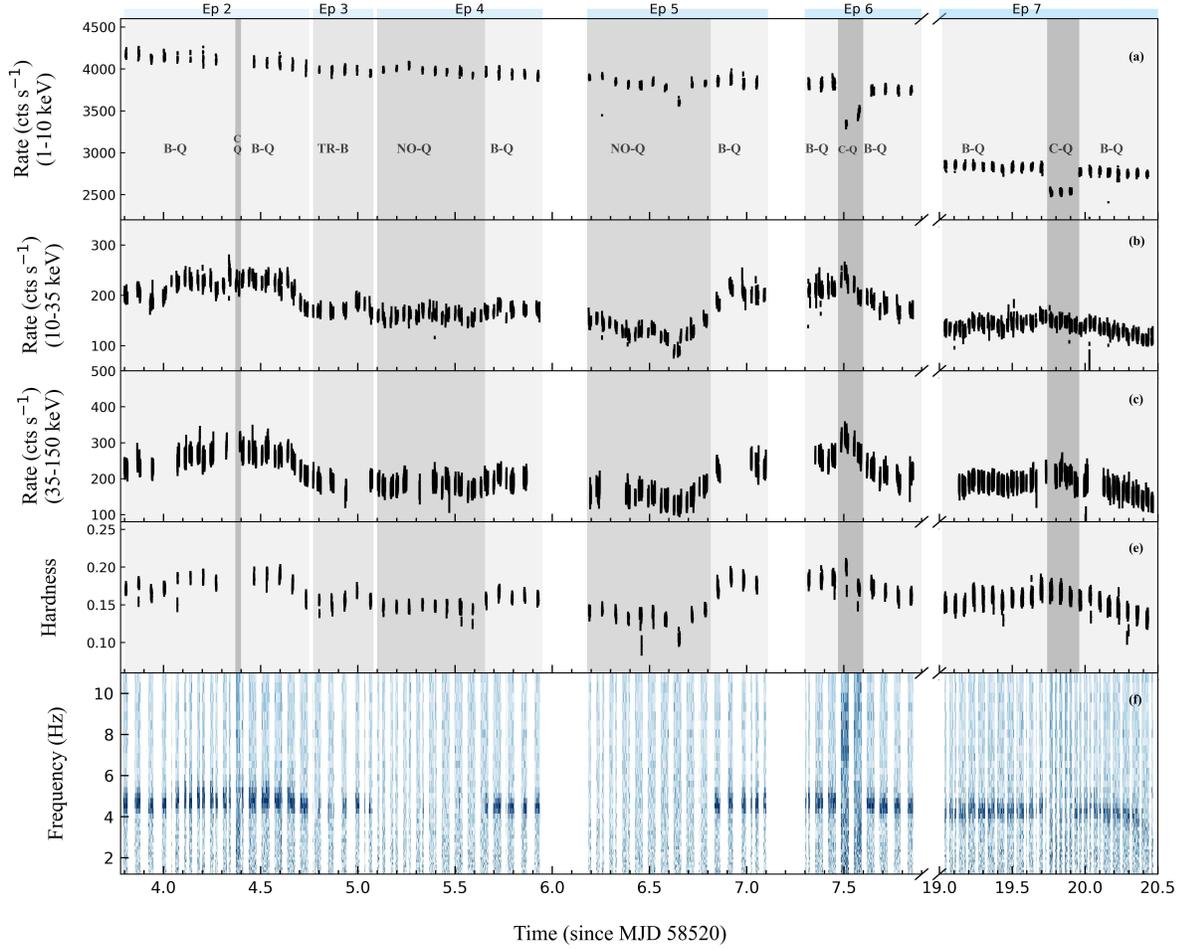}
    \caption{The detailed light curve and hardness (16 s bin) of the parts of SIMS \uppercase\expandafter{\romannumeral1} and SIMS \uppercase\expandafter{\romannumeral2} for this outburst. \emph{Panel (a), (b) and (c)}: light curves in three energy bands respectively: 1--10 keV (LE), 10--35 keV (ME) and 35--100 keV (HE). \emph{Panel (d)}: hardness between 4--10 keV (LE) and 1--4 keV (LE) count rate. \emph{Panel (e)}: dynamical power spectra (DPS) in the energy band 35--100 keV (HE) with a time resolution of 16 s. We separate the all SIMS into thirteen epochs. The label `C-Q' means type-C QPO are detected. Similarly, `B-Q' corresponds to type-B QPO, `NO-Q' corresponds to NO-QPO , `TR-B' corresponds to the time interval when the transitions of type-B QPO are detected.}
    \label{fig:Epoch}
\end{figure*}

In order to search for the QPOs, the PDS is normalized according to Leahy normalization \citep{1983ApJ...272..256L}. 
In Figure \ref{fig:Epoch}, we show the light curves from three energy bands (LE:1--10keV; ME:10--35 keV; HE: 35--100 keV), hardness and DPS with a time interval of 16 s for part of SIMS \uppercase\expandafter{\romannumeral1} (from MJD 58523.8 to MJD 58527.8) and SIMS \uppercase\expandafter{\romannumeral2} (MJD 58539.1--MJD 58540.2). The results show that in the bottom panel of Figure \ref{fig:Epoch}, type-B QPOs often disappear or transit into other types of QPOs when the counts rate change.

Based on the occurrences of QPOs and variations of rate shown in Figure \ref{fig:Epoch}, we divide the whole SIMS into 7 epochs and mark them by different colors, which are also referred in the last column of Table \ref{QPO}. Epoch 1 is the period from MJD 58522.60 to MJD 58524.25, during which no intermittent QPO and no significant changes in flux are observed. Therefore it is not shown in Figure \ref{fig:Epoch}. During Epoch 2 (MJD 58524.27--58524.41), a fast transition between type-B and type-C QPOs is detected. The fast re-occurrence of type-B QPOs on short time-scale is found during Epoch 3 (MJD 58524.77--58524.94). For Epoch 4 (MJD 58525.06--58525.87) and Epoch 5 (MJD 58526.19--58527.10),there are long time transition intervals between the appearance and disappearance of type-B QPOs. To clarify, the corresponding DPS are plotted in Figures \ref{fig:tr-b} and \ref{fig:tr-c}. In Epoch 6 and Epoch 7, the transitions between type-B and type-C QPOs are observed, accompanied by significant changes in the X-ray flux (also see Figure \ref{fig:tr-c}). 

Based on the DPS, the average PDS is produced according to Miyamoto normalization \citep{1991ApJ...383..784M} in units of (rms/mean)$^{2}$Hz$^{-1}$ after subtracting the Possion noise, in order to get the characteristics of QPOs for the different ObsIDs. 
The PDSs are fitted by a model consisting of multiple Lorentzians \citep{2002ApJ...572..392B}. The background contribution to the QPO fractional rms are corrected according to  \citep{1990A&A...227L..33B,2015ApJ...799....2B}
\begin{equation}
   {\rm rms} = \sqrt{R}\times \frac{(S+B)}{S},	\label{eq:quadratic}
\end{equation}
here \emph{R} is the power calculated with the integration of the QPO Lorentzian function over the frequency range 0.01--64 Hz, \emph{S} is the source count rate, while \emph{B} is the background count rate. We fit the PDS of all QPOs and the results are given in Table \ref{QPO}. The frequency of type-B QPOs varied within a narrow range of 4.3--4.7 Hz and 4.0--4.2 Hz in SIMS \uppercase\expandafter{\romannumeral1}
 (MJD 58522.7--58527.8) and SIMS \uppercase\expandafter{\romannumeral2}
 (MJD 58539.1--58540.2), respectively.

\begin{figure}
   \centering
    \includegraphics[scale=0.5]{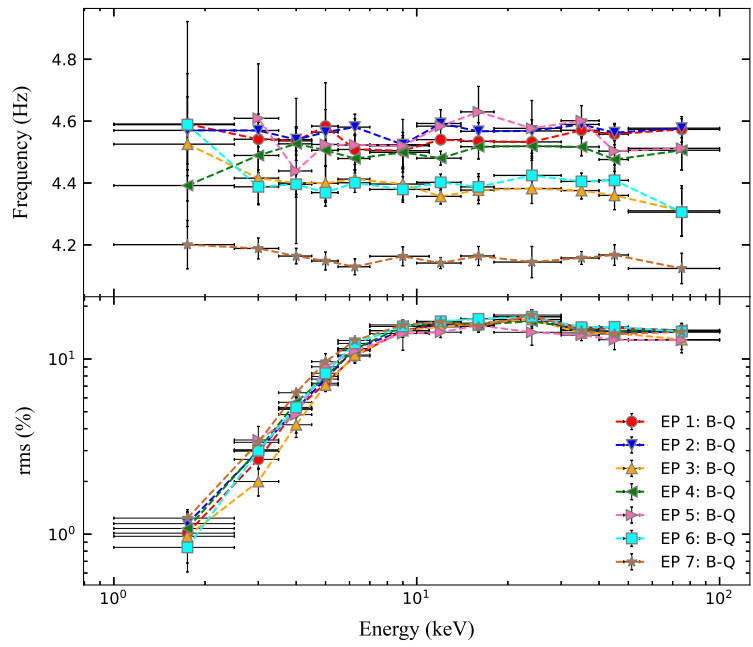}
    \caption{The relationship between type-B QPOs frequency/fractional rms and energy.}
\label{fig:qpo}
\end{figure}

In order to investigate the energy dependence of the type-B QPOs, type-B QPOs observations are divided into 12 energy bands from 1--100 keV: 1--2.5, 2.5--3.5, 4.5--5.5, 5.5--7.0, 7.0--10.0, 10.0--14.0, 14.0--18.0 18.0--30.0, 30.0--40.0, 40.0--50.0 and 50.0--100.0 keV. The corresponding centroid frequencies and the rms of type-B QPOs from all epochs are plotted in Figure \ref{fig:qpo}. The results show that rms increased with energy from 1 keV and reached a plateau above 10 keV. Meanwhile, the frequency of QPO remains unchanged with the energy.

\subsubsection{The temporary characteristics of Type-B QPOs}

\begin{figure*}
    \gridline{\fig{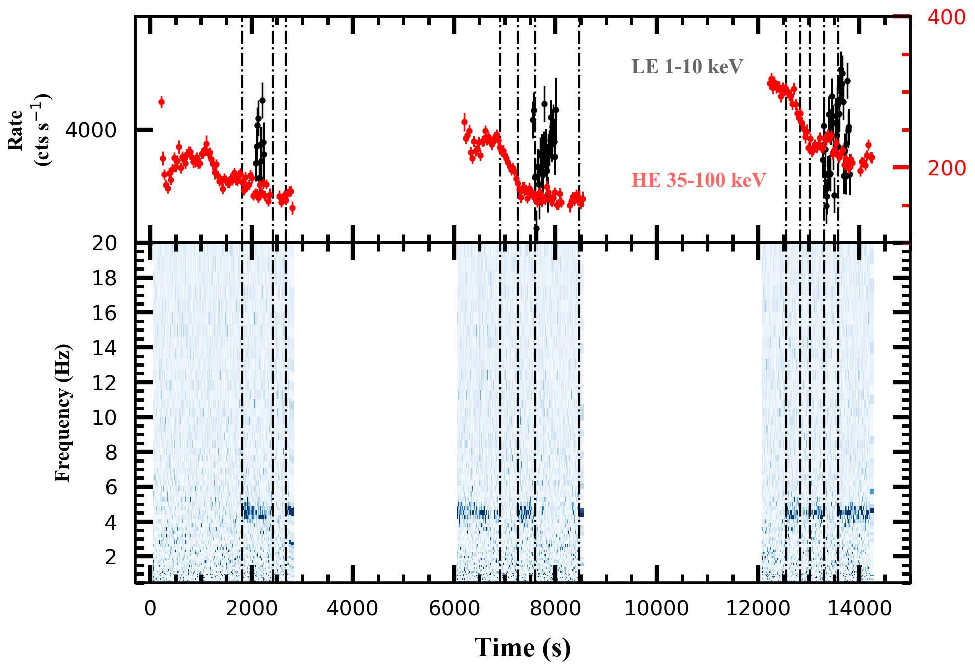}{0.45\textwidth}{}\fig{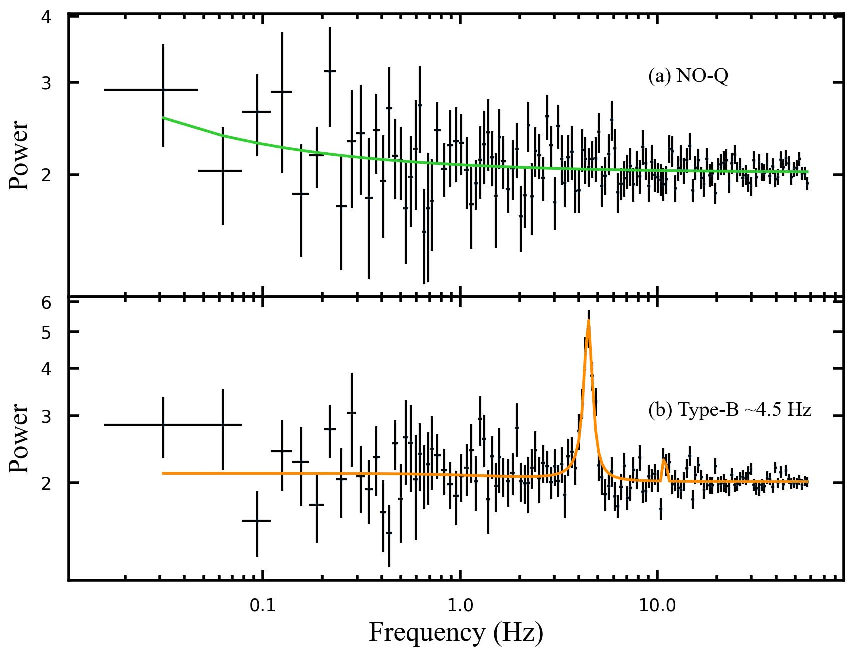}{0.45\textwidth}{}
}\vspace{-1.0cm}
    \gridline{\fig{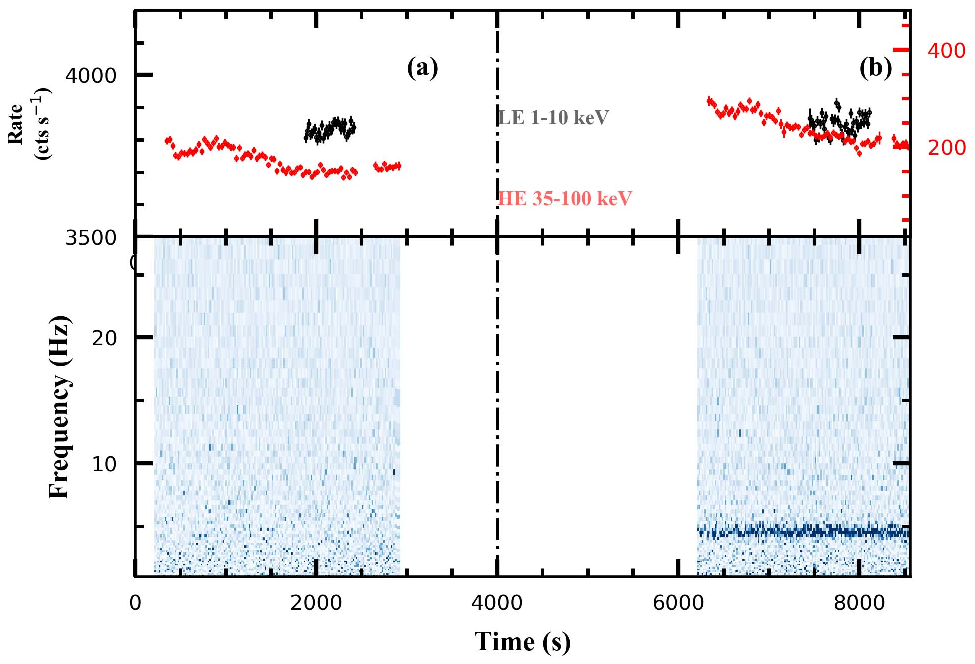}{0.45\textwidth}{}\fig{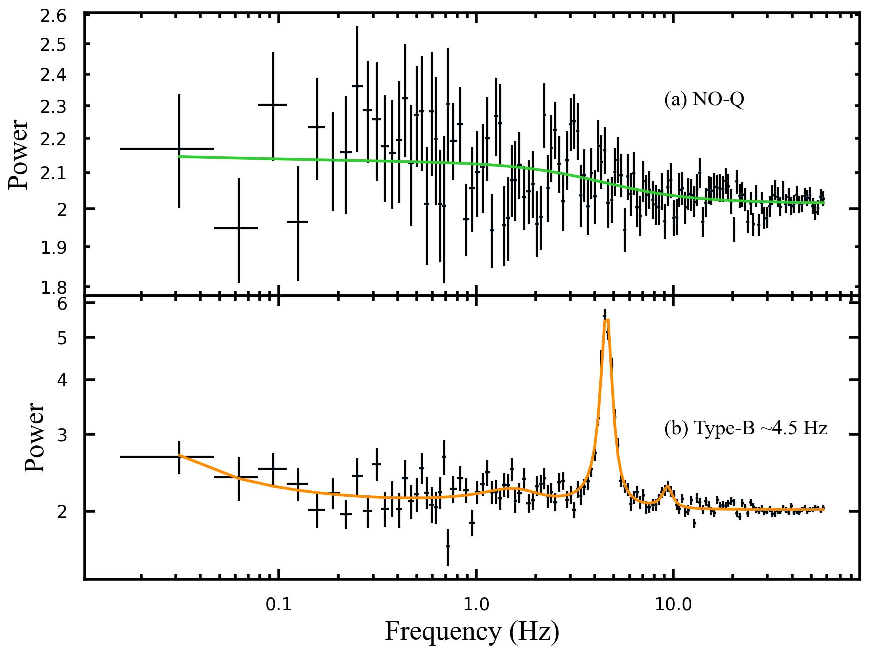}{0.45\textwidth}{}
}\vspace{-1.0cm}
    \caption{The detailed light curves (16 s bin), dynamical power spectra (DPS) of transition between NO-QPO and type-B QPOs and the corresponding average PDSs are shown on the right. The black and red lines represents the light curve of two energy bands: 1--10 keV (ME), 35--100 keV (HE). \emph{Top panel}: Fast transitions in Epoch 3 and the start time is MJD 58524.779. \emph{Bottom panel}: An example of long-time scale transitions involving  B-Q and NO-Q. The start time is MJD 58526.763 (Epoch 5). }
    \label{fig:tr-b}
\end{figure*}

In this work, we perform spectral analysis for these observations using two different models mentioned above in the energy band 1--100 keV. When source was in its SIMS, the QPOs frequently appeared or disappeared as shown in Figure \ref{fig:Epoch}. When the source entered Epoch 3 from Epoch 2, the count rates of ME, HE and the hardness decrease abruptly, while the DPS show that there are multiple rapid transitions between no QPO (hereafter NO-Q) and type-B QPOs (hereafter B-Q, also see in Table \ref{QPO}). In order to show the temporary nature of the type-B QPOs clearly, the DPS and PDS of partial Epoch 3 observations is shown in the top panel of Figure \ref{fig:tr-b}. We can see that type-B QPOs disappeared suddenly on a shorter time scale (about ten seconds), while no significant changes are observed in the flux. The PDS of NO-Q is characterized by a weak power-law noise. Unfortunately, the LE data is not available during this period, but previous studies using \emph{RXTE} and \emph{NICER}'s data have shown that the disappearance of type-B QPOs are often accompanied by a flux decrease in low energy band \citep{2013ApJ...775...28S,2021ApJ...911..127S,2021MNRAS.505.3823Z}.

Type-B QPOs can also be present or disappear for a few hours. In Epoch 1, type-B QPO are observed continuously for more than 3 days. In Epochs 4 and 5, QPOs are more likely to disappear completely at the relatively lower hardness and flux of the high energy bands (i.e. ME, HE), as shown in Figure \ref{fig:Epoch} (MJD 58525.1-58525.7 and MJD 58526-58526.8), which is consistent with the results of \citet{2021MNRAS.505.3823Z}. In the bottom panels of Figure \ref{fig:tr-b}, we show examples of DPS and PDS for a long-time scale transition between B-Q and NO-Q and find that the PDS of NO-Q show a weak power-law noise at low frequencies. We will discuss the possible physical causes of this transition in combination with information on time variation and energy spectrum in section \ref{Spectral Results}.

\subsubsection{Transitions between type-B and type-C QPOs}
\begin{figure*}
 \gridline{\fig{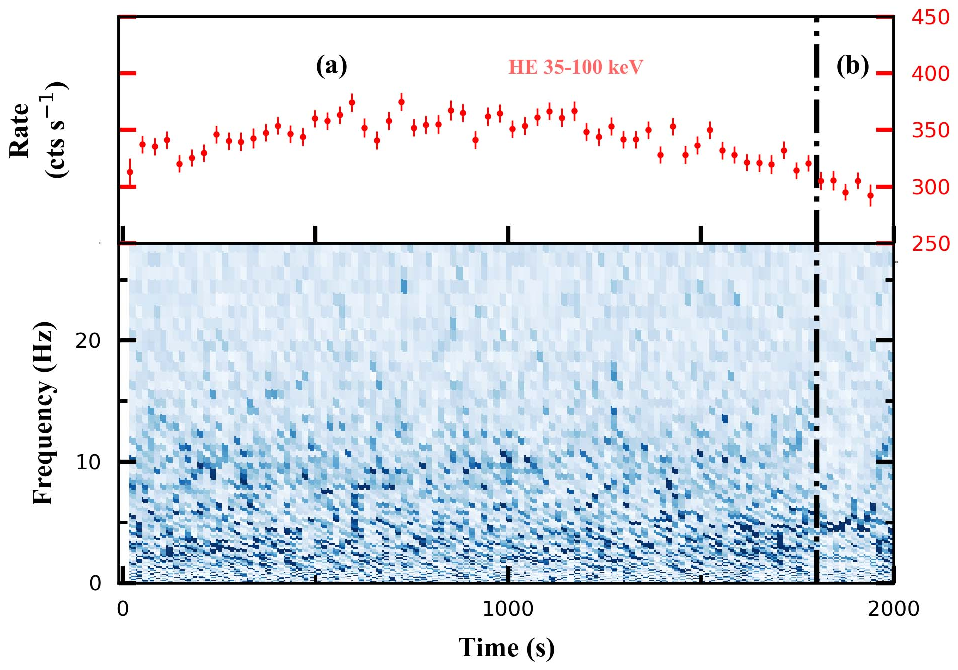}{0.45\textwidth}{}\fig{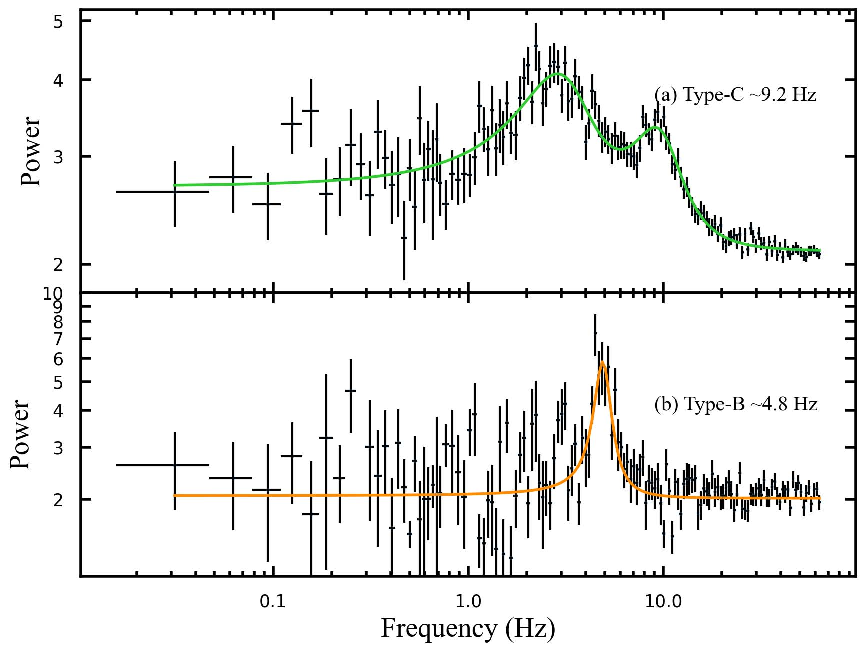}{0.45\textwidth}{}
}\vspace{-1.0cm}
\gridline{\fig{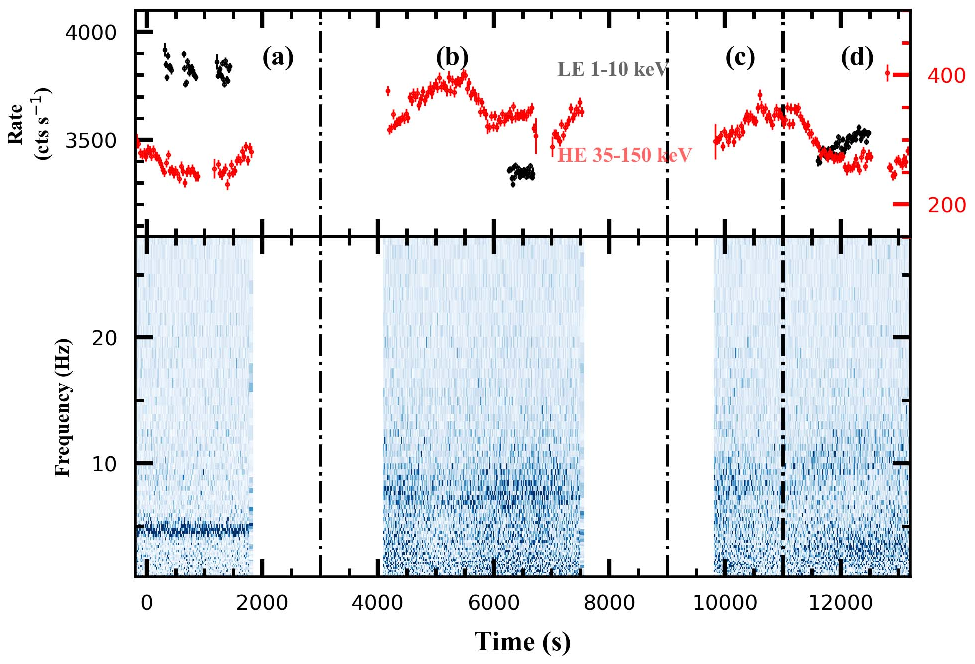}{0.45\textwidth}{}\fig{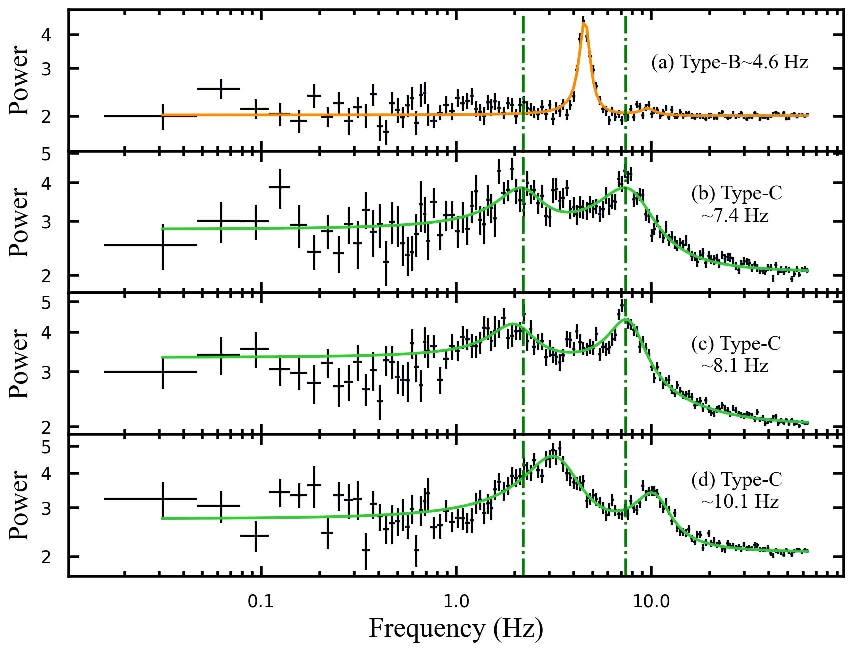}{0.45\textwidth}{}
          }\vspace{-1.0cm}
\gridline{\fig{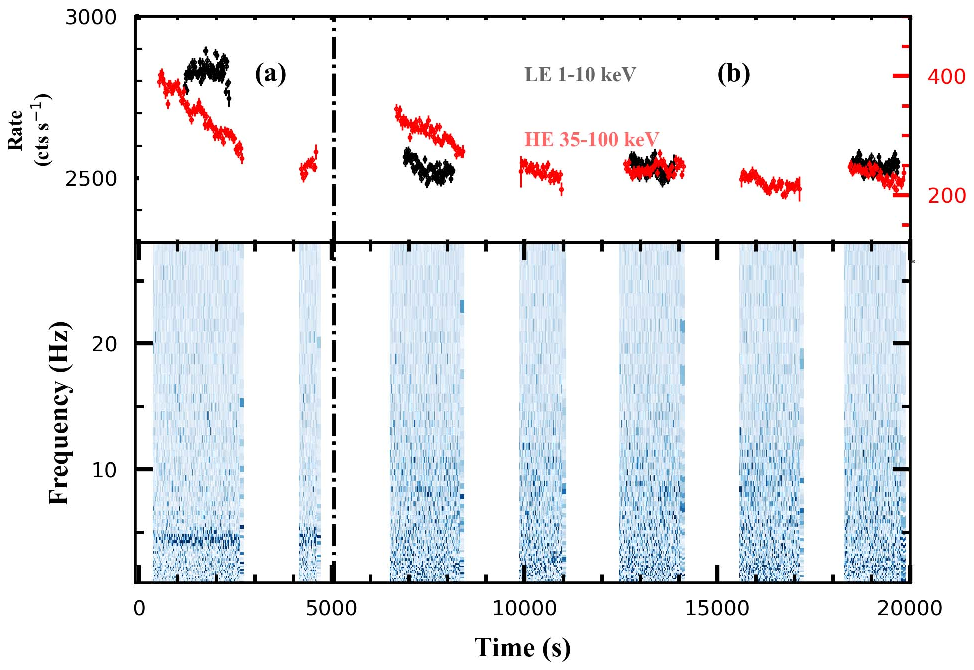}{0.45\textwidth}{}\fig{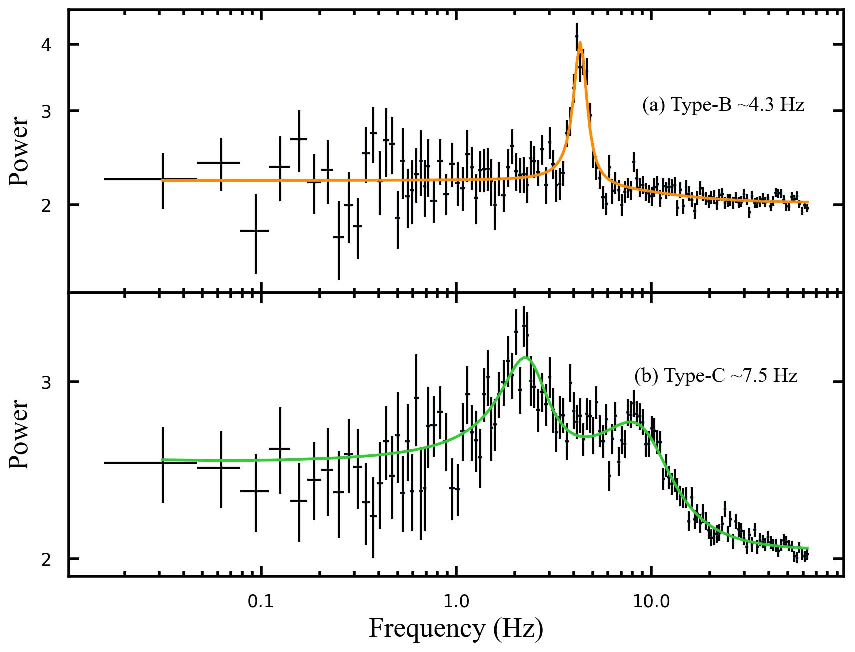}{0.45\textwidth}{}
          }\vspace{-0.7cm}
\caption{The detailed light curves and dynamical power spectra (DPS) (16 s bin) of the transitions between type-B and type-C QPOs are shown on the left. The black and red lines represents the light curve of two energy bands: 1--10 keV (ME), 35--100 keV (HE). The corresponding average PDSs are shown on the right. \emph{Top panel}: The transition between a type-B QPO and a type-C QPO in Epoch 2. The start time is MJD 58524.372. \emph{Middle panel}: Epoch 6 and the start time is MJD 58527.44. \emph{Bottom panel}: Epoch 7 and the start time is MJD 58539.75}
\label{fig:tr-c}
\end{figure*}

Several obvious transitions between type-B and type-C QPOs are observed in this outburst, as shown in Figure \ref{fig:tr-c}. The first one is observed on $\sim$ MJD 58524.4 (in Epoch 2), which occurred in a short period of $\sim$ 2000 s. During the first $\sim$ 1800 s, the averaged PDS show appearances of a QPO with two peaks of $\sim$ 2.8 Hz (rms $\sim$ 8.6\%) and $\sim$ 9.3 Hz (rms $\sim$ 10.3\%). Depending on the characteristics of type-C QPOs ($\sim$ 7 Hz with small rms $\leq$ 10\% in Huang et al.(In preparation)) at the HIMS of this source, it can be inferred that the QPO with frequency of $\sim$ 9.3 Hz is type-C. A type-B QPO with central frequency of $\sim$ 4.8 Hz (rms $\sim$ 8.1\%) is detected in the last $\sim 200$ seconds. It is likely that a type-C QPO appeared subsequently, though this is hard to confirm because of its short duration. This transition occurred in a particularly short period of time as we can see from the DPS shown in the top panel of Figure \ref{fig:tr-c}. 

The second transition (denoted as Epoch 6) is accompanied by a significant decrease in the count rate of LE and increase in the count rates of ME and HE (see the Figure \ref{fig:Epoch} and the middle panel of Figure \ref{fig:tr-c}), when a significant evolution of this type-C QPO is observed. The middle panel of Figure \ref{fig:tr-c} demonstrates that this QPO has two peaks and evolves over time; in particular, the lower frequency peak evolved from $\sim$ 2.9 Hz to $\sim$ 3.2 with a slight change in rms from 8.1\% to 8.2\%, while the higher peak evolved from 7.6 Hz to 10.0 Hz (rms $\sim$ 10.5\%-11.2\%).

During the SIMS \uppercase\expandafter{\romannumeral2}, the transition of a type-B at frequency of $\sim$ 4.3 Hz to a type-C QPO at frequency of $\sim$ 7.8 Hz is detected (denoted as Epoch 7), accompanied by a smaller low-energy rate (1--10 keV) and a larger high-energy rate ($\geq$ 10 keV), as shown in the bottom panel of Figure \ref{fig:tr-c}. The last two type-C QPOs also exhibited energy-dependent properties, i.e., the PDS exhibit different shapes for different energy bands.
\subsection{Spectral Analysis and Results}
\label{Spectral Results}
\begin{figure}
     \centering
     \includegraphics[scale=0.3]{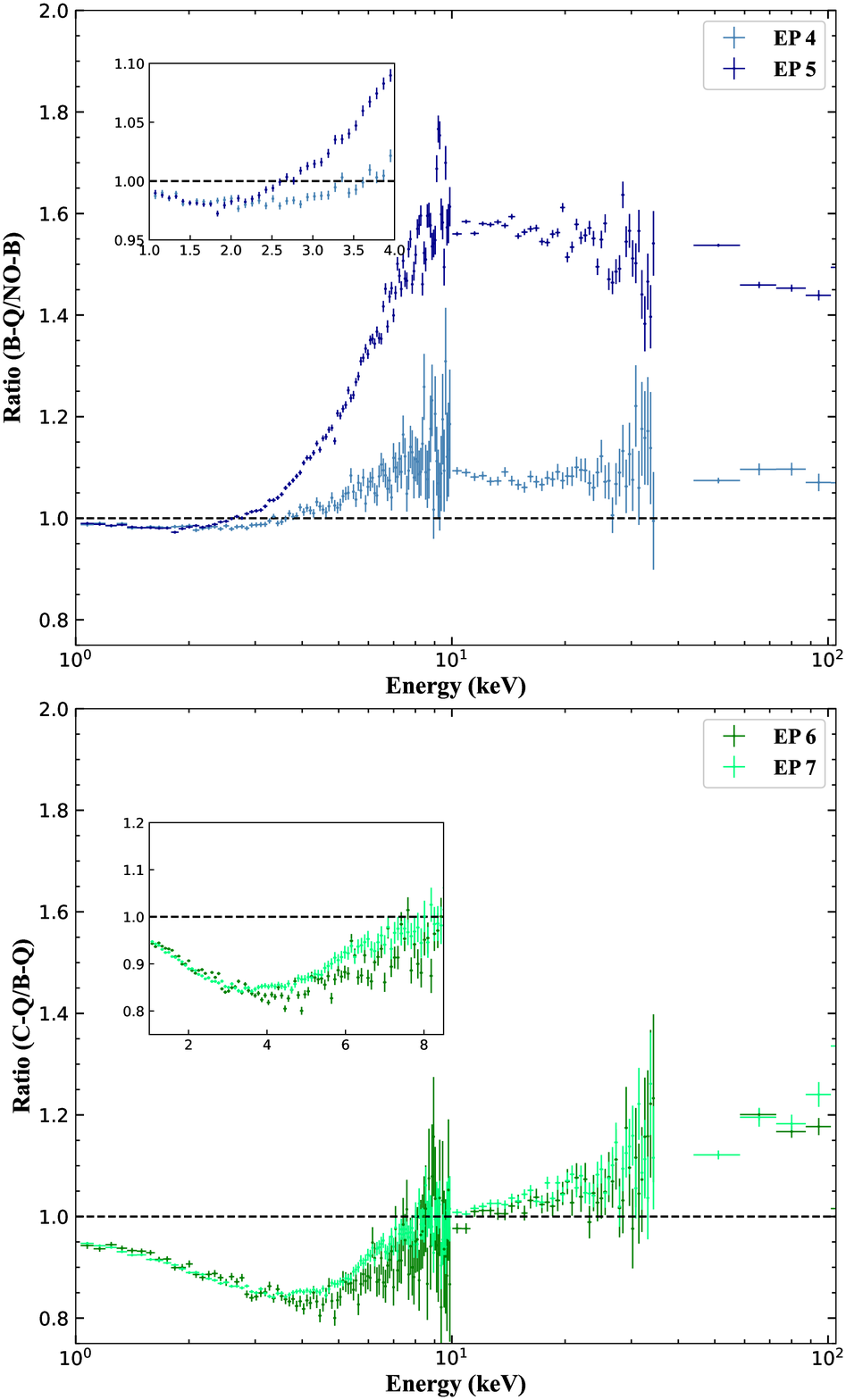}\vspace{-0.7cm}
     \caption{\emph{Top panel:} The spectral ratio defined as the ratio between the merged spectrum with the type-B QPO and that without the QPO (B/NO-Q), plotted in light-blue for Epoch 4 and dark-blue for Epoch 5. \emph{Bottom panel:} The spectral ratio defined as the ratio between the merged spectrum with the type-B QPO and type-C (C/B), plotted in green for Epoch 6 and light-green for the Epoch 7.}
     \label{fig:spectra ratio}
\end{figure}

Due to the limitation of LE GTIs, only data from Epoch 4 to Epoch 7 are used for the energy spectrum study. In order to study the energy spectral characteristics regarding different types of QPOs, we first combine the energy spectra based on PDS (i.e., NO-QPO, type-B QPO, and type-C QPO) that are observed in Epochs 4, 5, 6, and 7, namely NO-Q, B-Q, and C-Q, to increase the signal-to-noise. Before fitting the spectra, we first compare the spectra between the segments of NO-Q (C-Q) and B-Q qualitatively. The energy spectrum ratio between the B-Q and NO-Q observations is shown in the top panel of Figure \ref{fig:spectra ratio}. The ratios between B-Q and NO-Q indicate differences at both low and high energy bands. At energies below 3 keV, the ratio approaches 1, which indicates that the energy spectrum shape has slightly changed. However, the ratio is more or less constant above 10 keV, and the flux of B-Q is significantly higher than that of NO-Q at energies above 3 keV. These results are consistent with the phenomenons that type-B QPOs appear in the high energy bands. 

The energy spectra ratio between the type-C QPOs and type-B QPOs is shown in the bottom panel of Figure \ref{fig:spectra ratio}. We find that even after the source re-entered the SIMS \uppercase\expandafter{\romannumeral2} for 12 days, the trend of the energy spectra ratio remains the same when the QPO transitions occurred. The ratio (C-Q/B-Q) is appreciably less than 1 at energies \textless 10 keV (maximum value $\sim$ 0.8 at 4--5 keV), which indicates that the spectral shapes have significantly changed at low energy bands. At energies $>$ 10 keV, the ratio increases with energy with a value $>1$, which indicates that the C-Q flux is larger at higher energies. These agree with the fact that the spectra of type-B QPOs are softer than those of type-C QPOs in a transition. 

\begin{figure}
    \centering
    \includegraphics[scale=0.35]{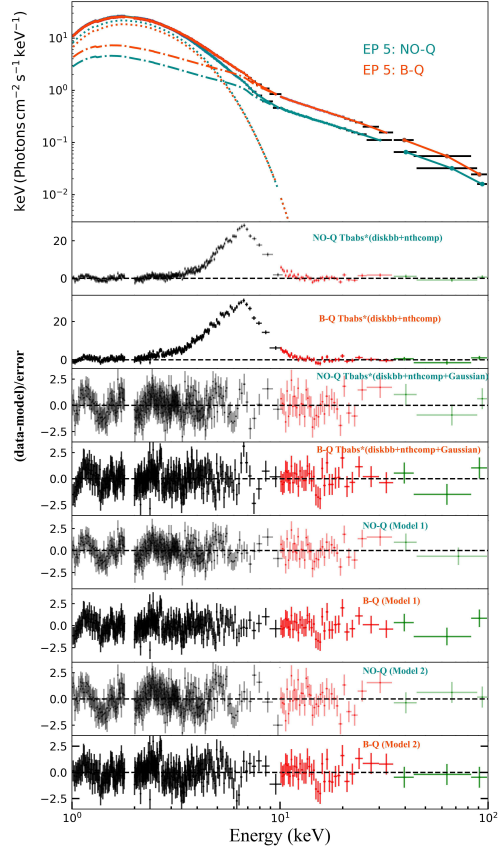}
    \caption{An example (The NO-Q and B-Q in Epoch 5) of fitting energy spectrum with four models. \emph{Top panel:} The total model (solid lines) is plotted together with contributions from the thermal disk (dotted lines) and the reflection component (dash-dot lines). \emph{The rest panels:} Spectral residuals from {\tt\string tbabs}$\times$({\tt\string diskbb+nthcomp}), {\tt\string tbabs}$\times$({\tt\string diskbb+gaussian+nthcomp}), {\tt\string tbabs}$\times$({\tt\string diskbb+gaussian+nthcomp+xillverCp}) as Model 1, and Model 2, {\tt\string tbabs}$\times$({\tt\string diskbb+relxilllp}). The spectra are plotted in dark-orange and dark-cyan for the B-Q and NO-Q section, respectively. Black, red, and green represent the spectra of the three detectors LE, ME, and HE, respectively  }
    \label{fig:example-spectra}
\end{figure}

To quantitatively study the evolution of spectral parameters, different models are applied to the observations in Epoch 5. As shown in Figure \ref{fig:example-spectra}, the first trial of fitting is with a model consisting of a multi-color disk component and a Compton component \citep{1996MNRAS.283..193Z, 1999MNRAS.309..561Z}, i.e., {\tt\string tbabs}$\times$({\tt\string diskbb+nthcomp}). The {\tt\string nthcomp} is the thermally comptonized continuum model that may arise from a hot corona associated with accretion disc in the BHTs. Therefore, during the fitting, we link the seed photon temperature ($T_{\rm bb}$) of the {\tt\string nthcomp} model with the inner disc temperature ($kT_{\rm in}$) of the $diskbb$ component. The goodness of fit ($\delta\chi^{2} > 2$) are mostly attributed to an obvious asymmetry of the broadened iron line. Accordingly, we add an extra {\tt\string gaussian} component to fit the spectra and the residuals are shown in the 4-5th panels of Figure \ref{fig:example-spectra}. However, a hint of a narrow iron line seems to be present in the residuals. Therefore, we add another non-relativistic reflection model {\tt\string xillverCp} with the inclination angle fixed at $36.5^{\circ}$, $\Gamma$ and $kT_{\rm e}$ linked to the parameters of {\tt\string nthcomp} and we calculate the significance using $F-{\rm test}$ in XPSEC.
The best-fitting parameters of model 1, i.e, {\tt\string tbabs}$\times$({\tt\string diskbb+gaussian+nthcomp+xillverCp}), are shown in Tables \ref{B-nthcomp} and \ref{C-nthcomp}. Although the fitting is significantly improved, the centroid line energy of Gaussian is less than the typical 6.4--7.1 keV iron regime, and the profile is extremely wide (1--2 keV). The values are un-physical, which is similar to the result in Cygnus~X-1 reported by \citet{2014ApJ...780...78T}. The broad emission feature is normally seen in the reflection spectra. 

\begin{figure*}
    \centering
  \includegraphics[scale=0.23]{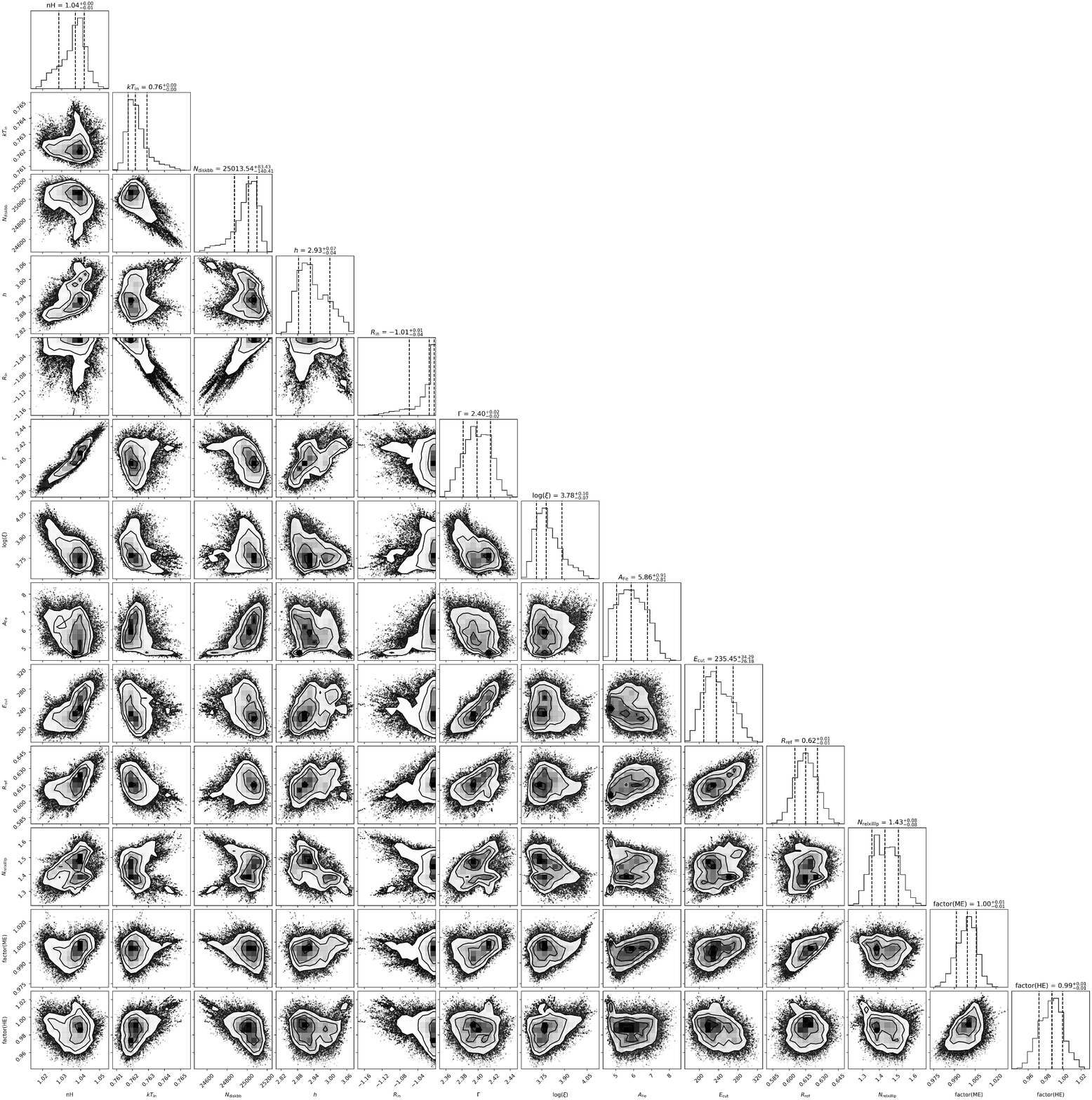}
    \caption{An illustration of one and two dimensional projections of the posterior probability distributions derived from the MCMC analysis for the parameters in model 2, {\tt\string tbabs}$\times$({\tt\string diskbb+relxilllp}). This illustration corresponds to the spectral fitting of NO-Q in Epoch 4.}
    \label{fig:mcmc}
\end{figure*}

An accretion disk with a lamppost geometry corona (the relativistic reflection model), as model 2, {\tt\string tbabs}$\times$({\tt\string diskbb+relxilllp}) is introduced to fit the spectra. For the {\tt\string relxilllp} components of all spectra, an extreme spin value of 0.8 and an inclination angle of $36.5^\circ$ (given in \citealt{2022arXiv220101207J}) are chosen and fixed during spectral fittings. We show the best-fitting parameters in Tables \ref{B-relxilllp} and \ref{C-relxilllp}. In the {\tt\string relxilllp} model, the reflection fraction $R_{\rm f}$ is defined as the ratio of the incident photon intensity that illuminates the accretion disc to that observed directly, $\Gamma$ and $E_{\rm cut}$ denote the initial spectral index and its deviation energy from a simple power-law shape. Other parameters in {\tt\string relxilllp} provide information about the accretion disk: the inner radius of disk $R_{\rm in}$, ionization of the accretion disk log($\xi$) and iron abundance of the material $A_{\rm Fe}$. However, the iron abundance is very high and often pegs at its limit during our fitting, probably because the reflection comes from a high-density disk, which can be referred to \citet{2021MNRAS.508..475C} for specific studies. An example (NO-Q in Epoch 4) of the MCMC (Markov Chain Monte Carlo) analysis for this model is present in Figure \ref{fig:mcmc}.

\subsubsection{The spectral results of B-Q and NO-Q}
To study the recurrences of type-B QPOs, we compare the segment NO-Q and B-Q energy spectra in Epochs 4 and 5 (also see Table \ref{fig:Epoch}). 
By using model 1, {\tt\string tbabs}$\times$({\tt\string diskbb+gaussian+nthcomp+xillverCp}), to fit the spectra in Epoch 4 and Epoch 5, we find that the disk temperature remains almost the same at $kT_{\rm in} \sim$ 0.74--0.75 keV along with the power-law index at $\Gamma \sim$ 2.20; the change in the normalization of {\tt\string diskbb} is also negligible within the error range. 

However, compared to NO-Q, the thermal flux is lower and the Comptonization flux is higher in B-Q. Especially for Epoch 5, the thermal flux is reduced by 5.3\% and the Comptonization flux increases by 58.3\%, which coincides with the direct energy spectrum comparison in Figure \ref{fig:spectra ratio}. To investigate which spectral parameter is responsible for the spectral changes, we perform a simultaneous fit, using NO-Q and B-Q spectra following the method of \citet{2013ApJ...775...28S}. First we fix the adjacent NO-Q and B-Q spectral parameters together and free them one by one to obtain the corresponding residuals and the changes of the parameters. For each epoch, the values of the segment B-Q spectrum are set to the best-fit parameters of segment NO-Q spectrum and the $\chi^{2}/d.o.f$ can be seen in the Table \ref{compare-nthcomp}. For Epochs 4 and 5, we free the {\tt\string nthcomp}, {\tt\string Gaussian}, and {\tt\string diskbb} normalization parameter and find that the fitting are improved significantly. Therefore, it can be assumed that some changes occurred in the corona and disk at this time.

In the same way as in model 1, the smallest set of parameters is listed in Table \ref{B-relxilllp} for Epoch 4 and Epoch 5. A relatively small variation in $N_{\rm diskbb}$ is observed in Epoch 4, whereas in Epoch 5 both $N_{\rm diskbb}$ and $kT_{\rm in}$ are found to be variable. Compared to the spectral results using the simple model, little change in photon index $\Gamma$ are observed for NO-Q/B-Q during this fitting. During the fitting of the four spectra, the corona height $h$ seems to be larger, while the change in $R_{\rm in}$ is insignificant $\sim 1 R_{\rm ISCO}$ (Radius of the innermost stable circular orbit). The inner disk radius can also be calculated from disk normalization parameter as outlined by \citet{1984PASJ...36..741M}, whereby $R_{\rm in}^{2} = f^{4}D_{10}^{2}(N_{\rm diskbb}/{\rm cos}\theta)$, where $f$ is the color correction factor \citep{1998PASJ...50..667K}, $D_{10}$ is the distance to the source in units of 10 kpc, and $\theta$ is the disk inclination. We adopt the best-fit $N_{\rm diskbb}$ values with their uncertainties from Table \ref{B-relxilllp}. Assuming a canonical value for color correction of $f=1.5$, a fixed disk inclination of $36.5^{\circ}$ and a fixed distance of 2.2 kpc, $R_{\rm in}$ is found to be $87.4\pm0.3$ km, $85.1^{+0.4}_{-0.8}$ km, for NO-Q and B-Q, respectively, in Epoch 4, and $91.3\pm0.1$ km, $81.5\pm0.3$ km in Epoch 5. The inner disk radius increases slightly in NO-Q if the assumption of $R_{\rm in}^{2} \propto N_{\rm diskbb}$ is considered.


\subsubsection{The spectral results of B-Q and C-Q}
Intuitively, it can be seen from the bottom panel of Figure \ref{fig:spectra ratio} that the disk component of B-Q is remarkably stronger compared to that of C-Q. In order to understand the change of accretion disk configurations in SIMS (B-Q) and HIMS (C-Q), we fit the energy spectra of Epoch 6 and Epoch 7 with models 1 and 2, respectively. From the best fitting results of the two models (see Table \ref{C-nthcomp} and Table \ref{C-relxilllp}), it can be inferred that for both B-Q and C-Q epochs, the parameters of the disk have changed significantly.

For both model 1 and model 2, the C-Q spectra had a relatively lower disk temperature $kT_{\rm in}$ and a larger inner disk radius ($N_{\rm diskbb}$) compared to the B-Q spectral state, meanwhile $\Gamma$ is slightly smaller in C-Q, suggesting a harder spectra. Correspondingly, the thermal component flux is weaker and the Comptonized component is stronger in C-Q. 

For Epochs 6 and 7, after freeing  $kT_{\rm in}$, $\Gamma$ and $N_{\rm diskbb}$, the residuals are improved significantly (\ref{compare-nthcomp}). These results suggest that the accretion disk moved inward when the source entered the SIMS from the HIMS, i.e, B-Q has a smaller inner radius than C-Q. This is consistent with the standard accretion disk model \citep{1997ApJ...489..865E}.

Using model 2 (see Table \ref{C-relxilllp}), it is also suggested that the normalization of the {\tt\string diskbb} is larger at C-Q. Using the parameters in Table \ref{C-relxilllp}, $R_{\rm in}$ is found to be $77.1\pm0.7$ km, $83.7\pm0.9$ km, $82.9\pm0.6$ km, for B-Q, C-Q, and B-Q, respectively, in the SIMS \uppercase\expandafter{\romannumeral1}. After 12 days in the SIMS \uppercase\expandafter{\romannumeral2}, $R_{\rm in}$ is found to be $82.9\pm0.3$ km, $87.5\pm0.2$ km, $85.3\pm0.3$ km for B-Q, C-Q, and B-Q. However, $R_{\rm in}$ in {\tt\string relxilllp} of C-Q is found to be the same as that of B-Q: $\sim 1 R_{\rm ISCO}$ for Epochs 6 and 7. We also find that the height of the corona $h$ is slightly larger in B-Q. 
For both epochs (see Table \ref{compare-relxilllp}), after freeing the parameters $kT_{\rm in}$, the chi-squared value are improved significantly. However, for Epoch 7, all the parameters need to be free except for log($\xi$). Furthermore, an improvement of fitting for Epoch 6 is also present after freeing $h$, which indicates that the height of the corona indeed changed.


From the above results, we conclude that after the source  MAXI~J1348--630 entered the SIMS (B-Q), it may have returned to the HIMS (C-Q) again, or returned to the SIMS (SIMS \uppercase\expandafter{\romannumeral2}) after HSS, suggested by changes in PDS and energy spectrum.

\section{Summary and Discussions}
\label{discussion}
In this paper, we focus on studying the timing and spectra properties of the QPO transitions observed in MAXI~J1348--630 using data from \emph{Insight}-HXMT during its 2019 outburst. The results can be summarized as follows:
\begin{itemize}
    \item The type-B QPOs ($\sim$ 4.2 -- 4.7 Hz) in this source are recurrent, on timescales from seconds to hours;
    \item The variation amplitude (rms) of QPOs has a same relation with energy, i.e., rms increases with energy below 10 keV and remains constant above 10 keV to 100 keV.
    \item When type-B QPOs disappear
    \begin{itemize}
    \item the PDS show a weak power-law noise;
    \item the flux of the Comptonization component decreases, while the spectral index $\Gamma$ remains unchanged;
    \item the inner disk radius increases slightly when inferred from disk component ($N_{\rm diskbb}$);
    \item the corona height of type-B QPOs observation is larger than that of NO-Q when including an additional reflection model {\tt\string xillver} by need for the continuum fitting. 
    \end{itemize}
    \item Three transitions are found between the type-C and type-B QPOs with the following results:
    \begin{itemize}
    \item a transient timescale as short as $\sim$ 10 seconds;
    \item the Comptonization flux is higher and the spectral index is smaller in C-Q;
    \item the inner radius of the accretion disk is moderately larger for C-Q as inferred from the disk component, while it remains constant $\sim 1 R_{\rm ISCO}$ using reflection model;
    \item the corona height is larger in B-Q using the reflection model. 
    \end{itemize}

\end{itemize}
The above results suggest that the type-B and C QPOs are associated with different physical mechanisms within the inner region of the accretion disk. In the next sub-sections, we will discuss those results and possible physical mechanisms of the QPOs.

\subsection{LFQPOs and Models}
\label{LFQPOs}
When the last type-C QPO at $\sim 7.35$ Hz in the HIMS disappeared, a subsequent type-B QPO detection means that MAXI~J1348--630 entered SIMS.
The spectral fitting results show that throughout the SIMS, the inner radius of the disc was very close to the ISCO, which is consistent with the results of \citet{2022arXiv220111919Z}. The same results have been found in several other BHTs, such as H1743-322 \citep{2021ApJ...911..127S}, XTE~J1859+226 \citep{2013ApJ...775...28S}, etc. 

The model proposed for type-B QPOs is not as comprehensive as for type-C, but there is much evidence that they may have originated from a jet-like structure. The most convincing evidence is that type-B QPOs and jet ejections (or radio flares) are observed to occur close in time \citep{2009MNRAS.396.1370F,2019ApJ...883..198R,2020MNRAS.498.5772R,2020ApJ...891L..29H}. 
Recently, observations on MAXI~J1820+070 \citep{2020ApJ...891L..29H} have given strong evidence that there is a connection between the appearance of type-B QPOs and the launch of discrete jet ejections because the short time delay ($\sim$ 2-2.5 hr) between the transition and a strong radio flare. During the current outburst of MAXI~J1348--630, two radio flares, considered to be associated with the discrete ejection of the same radio knot \citep{2021MNRAS.504..444C}, were detected in MJD 58523.2 and MJD 58527.7, which is reported by \citet{2019ATel12497....1C} and \citet{2020MNRAS.tmpL.236C} respectively. 
The first flare was about 0.6$\sim$0.8 day from the start of the SIMS. About $\sim 0.1$ day before the start of the second one, we observe a sharp drop in the count rate of LE and a slight rise in the count rate of ME/HE, while the averaged-PDS shows type-C QPOs. A type-B QPO with frequency $\sim 4.5$ Hz appeared almost simultaneously with the second radio flare, which started at MJD 58527.5 and peaked with flux $252 \pm 13$ mJy at MJD 58527.66, suggesting a strong connection between the type-B QPOs and the ejection. 

For the type-B QPOs observed in this source, using the observations obtained with the \emph{NICER} observatory, \citet{2020MNRAS.496.4366B} computed the energy dependence of the rms and the phase lags at the QPO frequency, sampling for the first time at energies below 2 keV. Concentrating on the phase lags, \citet{2021MNRAS.501.3173G} successfully used a two-component Comptonization model to fit them and explained the radiative properties of the type-B QPOs in 0.8--10 keV energy range by two physically connected Comptonization regions.

Our result shows that the fractional rms of type-B QPOs increase with energy up to 10 keV and stay more or less constant above 10 keV in all observations, which implies a Comptonization origin of the type-B QPO.
In the jet precession model, the jet velocity can be measured by the QPO amplitude \citep{2021NatAs...5...94M}. Assuming the QPOs rms of $\sim 15\%$ (see Figure \ref{fig:qpo}; B-QPO rms), an observed inclination angle of $\sim 30^{\circ}$, and the jet half-opening angle of $\sim 3^{\circ}$, the jet velocity can be determined as $0.55-0.9c$. Assuming $v = \beta \times c$ to be the escape velocity of the black hole, the height of the ejection knot can be calculated as $\sim 6.6-2.5 r_{\rm g}$ ($\beta = 0.55-0.9 c$) , which is consistent with the emitting scale height of the hard component (Table \ref{B-relxilllp}). Recently \citet{2020A&A...640A..18M} also have found that type-B QPOs require a small transition radius ($r_{J}$, truncation radius due to a jet in the inner region) around 2.5 $r_{\rm g}$. These results indicate that type-B QPOs are produced very close to the BH. 

Furthermore, the similar rms--energy correlations for different LFQPOs in the 2-30 keV range are found with \emph{RXTE} in XTE~J1859+226 \citep{2004A&A...426..587C}, GRS~1915+015 \citep{2004sf2a.conf..437R,2012Ap&SS.337..137Y,2013ApJ...767...44Y,2016ApJ...833...27Y}, H1743--322 \citep{2013MNRAS.433..412L}, and XTE~J1550--564 \citep{2013MNRAS.428.1704L}, ), in which a corona origin of type-C QPOs is considered.
Thanks to the large effective area of \emph{Insight}-HXMT at high energies, the similar rms--energy correlations above 30 keV have been found in MAXI~J1535-571 \citep{2018ApJ...866..122H}, MAXI~J1820+070 \citep{2021NatAs...5...94M} and MAXI~J1631-479 \citep{2021ApJ...919...92B}. Their results all suggest a geometric origin of the type-C QPOs, while \citet{2021NatAs...5...94M} proposed that the type-C QPOs originate from jet rather than corona.

Therefore, whether the origin of type-C QPOs lies in the corona or the jet is still an open issue. A recent study in GRS 1915+105 \citep{2022NatAs.tmp...80M} proved that the X-ray corona can morph into a jet. They found that the corona is hot and extended, and covers the inner parts of the disk when the type-C QPO frequency is in the 2--6 Hz range. Then the corona will become a jet as the type-C QPO frequency decreases below $\sim 2$ Hz. A similar conclusion was also proposed by \citet{2022arXiv220511581A} for MAXI~J1348--630, who suggested that there are differences in the geometry of the system at the different phases of the outburst according to the relationship between QPOs frequency and their phase lags.

During the the SIMS in MAXI~J1348--630, three transitions from type-B QPO to type-C and then to type-B are observed by \emph{Insight}-HXMT. One of them is only detected by ME and HE, but it is a continuous transition from a type-C QPO at $\sim$ 9.2 Hz to a type-B QPO at $\sim$ 4.8 Hz in a short period of $\sim 10$ s (see the top panel of Figure \ref{fig:tr-c}), which also occurred in MAXI J1820+070 \citep{2020ApJ...891L..29H}. However, the flux at the high energy band changed very slightly in our observation. This may indicate that both the type-C and B QPOs are either generated in different regions located at different radii or in same region with different modulations.  

For the other two occurrences of type-C QPOs between type-B QPOs, we can present not only the differences in the power spectra but also the differences in the energy spectra. The bottom panel of Figure \ref{fig:spectra ratio} clearly shows that the type-C QPOs have harder spectra than type-B in energy above 10 keV, and less flux at low energies. Furthermore, it can be inferred from Tables \ref{C-nthcomp} and \ref{C-relxilllp} that the energy spectra of C-Q have less disc contribution and low disk temperature ($kT_{\rm in}$). The non-relativistic model seems to be less significant at C-Q in the fit of model 1, probably because of the contraction of the jet structure. This hypothesis is consistent with the $h$ parameter change in model 2. The inner disk radius of C-Q is moderately larger as inferred from the disk normalization parameter: $\sim1.71R_{\rm ISCO}$, $\sim1.86R_{\rm ISCO}$, $\sim1.84R_{\rm ISCO}$ for B-Q, C-Q, and B-Q, respectively, in the SIMS \uppercase\expandafter{\romannumeral1}. However, $R_{\rm in}$ given by the reflection model remains at $R_{\rm ISCO}$. One possible reason for this is that the broad iron line leads to a small radius when applying the reflection model, even in the hard state \citep{2010MNRAS.402..836R,2022ApJ...932...66R}. A recent simulation study reported by \citet{2022arXiv220103526L}, assuming cold, optically thick, clumps of gas falling into the black hole, also suggested that a broader iron line usually appears in the harder spectral states. Another reason could be that the inner radius varied so little during the transitions that the reflection model is insensitive to it. We test this by fixing $R_{\rm in}$ to the radius value derived from $N_{\rm diskbb}$ and find that the effect on the other parameters is insignificant.

We consider the L-T precession under the truncated disc/corona (hot inner flow) geometry \citep{2009MNRAS.397L.101I}, assuming that both QPOs $\sim$ 7--9 Hz and $\sim$ 4--5 Hz originate from the precession of an inner flow, a $\sim$ 7--9 Hz QPOs spectrum should have a smaller inner radius than the latter. The observations reveal the difference of quality factor ($Q$), suggesting that the precessions of different morphological structures leads to different $Q$. The type-B QPOs appear when the disk is close to the ISCO, which may indicate that the motion of the disk caused a deformation of the Comptonization region, and thus the flux modulation of different properties is observed in MAXI~J1348--630. \citet{2020A&A...640L..16K} quantitatively explained that the type-B QPOs in GX~339--4 come from the precession of non-solid structures, while type-C QPOs originate from the precession of solid inner flow. In the simulation results of \citet{2018MNRAS.474L..81L}, a solid-body-like precession of the tilted disk-jet system can explain type-C QPOs. Nevertheless, it is also difficult to explain the difference between type-C and B QPOs using solid/non-solid body behavior, which requires better simulations and higher quality observations.

Based on the above, we find that both type-B and C QPOs may be related to the so-called jet. Nevertheless, it is clear that types C and B QPOs have significantly different timing and spectral features as summarized in GX~339--4 by \citet{2011MNRAS.418.2292M} as well as in MAXI~J1348--630. Their common Comptonization origin properties and the short transition time scale suggest that a rapid change in physical properties or morphology of the Comptonization results in their differences. 

The jet/corona in low-luminosity states (hard states) is suggested to be formed in the BP mechanism (\citealt{1982MNRAS.199..883B}. see also e.g. \citealt{1999MNRAS.303L...1B,2001ApJ...548L...9M}). \citet{2001ApJ...548L...9M} further suggested that during very high accretion states, the more relativistic jet rather than the `BP' jet may be associated with the BZ mechanism \citep{1977MNRAS.179..433B}. This jet has a higher Lorentz factor, causing the propagation of an internal shock through the slower-moving outflow in front of it, which may lead the observation of discrete ejection \citet{2004MNRAS.355.1105F}.
Therefore, we argue that the compact and steady `BP' jet in low accretion phases is associated with type-C QPOs, while the fast `BZ' jet in SIMS is associated with type-B QPOs. As the accretion rate increases, the truncation radius moves inward, resulting in an increase in the frequency of type-C QPOs \citep{2006A&A...447..813F,2022arXiv220103526L}. Broadband noise in PDS during hard states are assumed to arise from the propagation of magneto-rotational instability (MRI; \citealt{1991ApJ...376..214B}) fluctuations of the inner hot flow \citep{2009MNRAS.397L.101I,2011MNRAS.415.2323I,2022ApJ...932....7Y}. 
During the SIMS phase, the frequency range of type-B QPOs varies narrowly, which may be related to a particular radius and a detailed discussion is given in sub-section \ref{Type-B QPO}. The noise weakness during SIMS may be due to the magnetic field of `BZ' jet emanating from the black hole, meanwhile the MRI originating with the disk is suppressed. 

\subsection{The disappearances of Type-B QPOs}
\label{Type-B QPO}
The disappearance of type-B QPOs (NO-Q/B-Q) has been studied using archival data from the \emph{RXTE}, for example, on XTE~J1817--330, XTE~J1859+226, and  H1743--322 \citep{2012A&A...541A...6S,2013ApJ...775...28S,2021ApJ...911..127S}. From the energy spectra comparisons (Figure \ref{fig:spectra ratio}), it can be seen that the difference is very small at energies below 2 keV where the emission is dominated by the thermal component. The most noticeable difference is in the energies above 2 keV, suggesting that the disappearance of type-B QPOs may be related to changes in the Comptonized component. The same result has been found in this source on the results of \citet{2021MNRAS.505.3823Z} and several of the above sources. The observed transient properties of type-B QPOs can occur on timescales of seconds to hours, which implies that the Comptonization radiation region is unstable. When type-B QPOs appeared, the increase in the high-energy flux indicates that there is either an extra Comptonized region contributing to the energy spectra, or the flux in the same region increases for some reasons. Our results give that the height of the corona is greater at B-Q, although the difference in the spectral shape is insignificant (small change in $\Gamma$). For the type-B QPOs observed in this source, using the observations obtained with the \emph{NICER} observatory, \citet{2021MNRAS.505.3823Z} gave a hint that an additional component involving the transitions of type-B QPOs may be related to the base of jet. 

In section \ref{LFQPOs}, we discussed that the type-B QPOs originating from the jet near the BH and the disk is also very close to the ISCO at this time. Therefore, we suggest that a coupled inner disk-jet region is responsible for the observed type-B QPOs transitions. 

Although the jets powered by BZ and BP effect are continuous and steady, the magnetic field lines close to the BH can be complicated by the strong gravity. However, the simulation results of different ways of \citet{2019MNRAS.484.4920Y,2019MNRAS.487.4114Y} and \citet{2021MNRAS.507..983L} allow us to understand the precessing jet near the BH and its instability from two perspectives. 

Theoretical studies show that the jet with transient properties can only exist when the accretion disk is very close to the BH and the magnetic fields on the accretion disk are quite in-homogeneous \citep{2022arXiv220100512Y,2021EGUGA..2316423L,2009RAA.....9..725W,2019MNRAS.484.4920Y,2019MNRAS.487.4114Y,2009MNRAS.395.2183Y}. The small-scale magnetic flux tube model proposed by \citet{2019MNRAS.484.4920Y,2019MNRAS.487.4114Y} argue that a closed zone with a size of a few gravitational radii is formed by the magnetic flux tubes, which may arise due to magnetorotational instability or magnetic buoyancy. This zone connecting the central compact object and the accretion disc may be able to produce strong X-ray emission and can be seen as the small-scale jet, which precesses and modulates X-ray emission (LFQPOs) as the BH rotates. Considering that the type-B QPOs arise from the precession of small-scale jets that are produced only when the disk is close to the BH, it is natural that the frequency of the type-B QPOs varies within a narrow range. Obviously, the field lines in and around this closed structure are actually continuously reconnecting i.e. the field lines connecting a point of the BH to a point on the disk only maintain the connectivity for a short time, and the rotation of the BH increases the toroidal twist of this field line, which eventually breaks. After that, open field lines are formed on the disk. As a result, the field lines on the accretion disk around the closed region alternate between closed and open configurations, leading to the sporadic type-B QPOs observed as well. The Comptonization emission still observed in NO-Q may be due to the remaining symmetric corona or a collimated smaller-scale jet (smaller height in NO-Q).

In addition to the small-scale magnetic tube model, another mechanism involving titled disks can explain the rapid transitions of type-B QPOs. Because the accreted material is most likely independent of the BH spin, its angular momentum is expected to be misaligned with respect to the BH spin, resulting in a titled accretion disc. \citet{2019MNRAS.487..550L} performed 3D general relativistic magnetohydrodynamics (GRMHD) simulations of tilted discs around rapidly spinning BHs. They suggested that such disk undergo L-T precession and can launch relativistic jets that propagate along the disc rotation axis. As we discussed above, the flux modulation produced by such jets that precess with discs allows us to observe type-B QPOs.  
\citet{1975ApJ...195L..65B} predicted that in the presence of realistic magnetized turbulence, the inner part of the disk ($< r_{\rm BP}$, Bardeen-Petterson radius) can align with the BH spin axis, while the outer disc is titled.
Recently, \citet{2021MNRAS.507..983L} found a smaller ($\lesssim 5-10 r_{\rm g}$ ) alignment radius than predicted by analytic models through GRMHD simulations, which is consistent with our energy spectral fitting results.

When the disk reaches the Bardeen-Petterson transition radius, the jet is parallel to the spin axis of the BH due to the dragging effect, and type-B QPOs generated by the jet precession will not appear. The observed PDS noise at this time is very small, probably due to the collimation of the jet. However, the Comptonization component is observed at this time, indicating that it is different from the HSS. Therefore, it is highly likely that there is a parallel jet generating Comptonization radiation, but without obvious modulations in its flux. The results show that the Comptonization flux decreases when the disk tears, probably because the outer disc cannot maintain a sufficient gas supply \citep{2021MNRAS.507..983L}. Conversely, disk flux less than 2 keV increases slightly, which is also seen in recent simulations of \citet{2022arXiv220103085M}, although their main research is not type-B QPOs. They suggested that the absence of a warp in the aligned portion of the disk reduces the inflow speed of the gas, which increases the density.
Furthermore, the obtained rms-spectra are more or less the same regardless of how much the flux of the hard component increases when type-B QPOs are present, suggesting that type-B QPOs are more likely to arise from the overall behavior of one component.

It is accepted that the transient jets (discrete ejection) on large-scales could be the result of the ejection of the corona at state transitions \citep{2003ApJ...595.1032R,2003ApJ...597.1023V} and are associated with presence of type-B QPOs. In addition, the radio flares in XRBs are proposed to be produced by the adiabatic expansion of ejected plasma blobs \citep{2004MNRAS.355.1105F,2012MNRAS.421..468M}; hence the ejection components should be launched before strong radio flares are observed. If the ejections of the corona (Comptonization part) are associated with the absence of type-B QPOs, although type-B QPOs occasionally disappear before or after ejections, it is possible that the opened magnetic field lines (parallel to the BH axis) can accelerate the plasma to a very large distance, leading to the in radio band discrete ejections. 

\subsection{Unified model for type-C and type-B QPOs}
More and more observational results in the BHT family indicate that LFQPOs may be a geometrical effect, which is also supported by our findings in MAXI~J1348--630. However, in the geometrical effect models, explaining the transition between different type QPOs, as well as the transient properties of type-B QPOs remain open issues. 

\begin{figure*}
    \centering
    \includegraphics[scale=0.4]{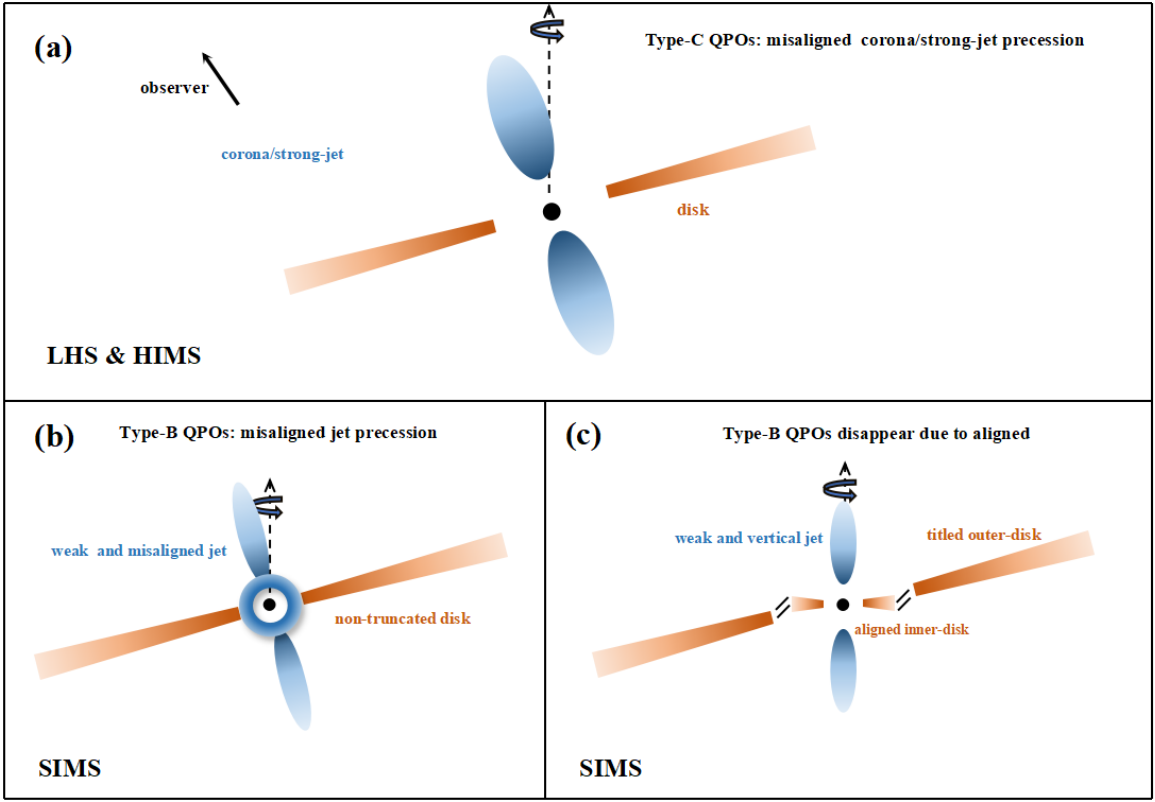}
    \caption{A sketch of the different states during an outburst. \emph{Panel (a)} illustrates the Comptonization component dominated states: LHS and HIMS, where type-C QPOs originate from the L-T precessing of the corona or jet. However, it cannot be determined if the disk is truncated. \emph{Panel (b) and (c)} show SIMS when type-B QPOs are present or absent and the radius of the disk is close to $R_{\rm ISCO}$. \emph{Panel (b)} depicts a jet in the same direction as the titled disc. Due to the flux modulation caused by the precession effect, we observe type-B QPOs. When the jet is parallel due to Bardeen-Petterson effect to the BH spin axis, the modulation effect on the flux disappears, as shown in \emph{panel (c)}, where the type-B QPO disappears.}
    \label{fig:model}
\end{figure*}

Based on the above results in BHT MAXI~J1348--630, we propose a scenario illustrated in Figure \ref{fig:model} to explain the timing and spectral evolution as observed in the BHTs. In the LHS and HIMS (panel (a) of Figure \ref{fig:model}), the L-T precession of the corona/jet in the inner regions of the disk (or titled disk) produces type-C QPOs due to the misalignment with BH spin axis. With the increase of accretion rate, the inner edge of the accretion disk gradually approaches the BH, while the corona/jet becomes smaller and its precession frequency increases. Although the jet-structure shown in top panel of Figure \ref{fig:model}, considering frequency ($\sim 7-10$) Hz and the variation in spectra of type-C QPO, we argue that it may be possible that the vertical-jet becomes a corona structure during the QPO transitions, or vice versa (B-C-B), as suggested in GRS 1915+105 \citep{2022NatAs.tmp...80M}.

In the SIMS, the inner radius of the accretion disk approaches its ISCO (panel (b) of Figure \ref{fig:model}) and the precessing jets parallel to the titled disc produce type-B QPOs. The innermost parts of the accretion disk align with the BH equator due to the Bardeen-Petterson effect, while the outer parts remain tilted and the disc forms a smooth warp in between (panel (c) of Figure \ref{fig:model}). Since the jets are launched along the direction of the BH spin axis, type-B QPOs disappear due to the absence of precession effect on the modulation of the flux. Alternatively, we can also explain the absence of type-B QPOs by the disappearance of small-scale jets bound by the magnetic flux tubes near the BH. In SIMS, possibly for some reason, inner disc may subsequently recede, when the larger scale Comptonisation component (corona/strong-jet) dominates the radiation and therefore the appearance of type-C QPOs may be observed. But what causes the outward movement of the disk is not yet precisely conclusive.

\section{Conclusions}
\label{conclusion}
In this work, we investigate the spectral and timing properties of the type-B QPOs transitions found in the BHT source MAXI~J1348--630. An additional {\tt\string xillver} component is required in the modelling of the continuum of type-B QPO, while the reflection modelling also suggests a larger height of the corona. Combined these with other results, we suggest that type-B QPOs probably originate from the precession of the weak-jet in a titled disk-jet structure located relatively close to the BH. However, due to the Bardeen-Petterson effect, the misaligned inner-disk in the high spin BH system can tend to be parallel to the direction of the BH spin axis, which can explain the disappearance of type-B QPOs in the BHXBs. We can also explain the absence of type-B QPOs by the disappearance of small-scale jets bound by the magnetic flux tubes. In this source, \emph{Insight}--HMXT observed a type-C/B transition that occurs within a very short time scale. It is still unclear whether type-C QPOs originate from the jet or the corona precession. Based on our results, there is only a slight increase in the inner disk radius when type-C QPO appears. However, due to the limitation of the observation time, it is not clear whether the disc is moving outward or the reasons for the type-C/B QPO transitions in SIMS.

\begin{acknowledgements}
This work made use of the data from the mission,a project funded by China National Space Administration (CNSA) and the Chinese Academy of Sciences (CAS).This work is supported by the National Key R\&D Program of China (2021YFA0718500). This work is supported by the National Natural Science Foundation of China (Grant No. 11733009, 11673023, U1838202, U1838201, U1938102, U2038104 and U2031205.), the CAS Pioneer Hundred Talent Program (grant No. Y8291130K2) and the Scientific and technological innovation project of IHEP (grant No. Y7515570U1).
\end{acknowledgements}

\bibliography{maxi}{}
\bibliographystyle{aasjournal}

\begin{deluxetable*}{cccccccc}[!htbp]
\label{QPO}
\tablecaption{QPO types classification and variability parameters.}
\tablewidth{500pt}
\tabletypesize{\scriptsize}
\tablehead{
\colhead{ObsID} & \colhead{Start-Time} & \colhead{End-Time} &
\colhead{QPO Type$^{a}$} & \colhead{Frequency$^{b}$} & 
\colhead{${{\rm FWHM}_{\rm QPO}}^{c}$} & \colhead{${{\rm rms}_{\rm QPO}}^{d}$} &
\colhead{Epoch$^{e}$}\\
\colhead{} & \colhead{MJD} & \colhead{MJD} &
\colhead{} & \colhead{(Hz)} & 
\colhead{(Hz)} & \colhead{(\%)} 
} 
\startdata
P021400201101 & 58522.607 & 58522.756 & B & $4.66 \pm 0.02$ & $0.54 \pm 0.05$ & $13.9 \pm 0.1$ & {}\\
P021400201102 & 58522.808 & 58522.889 & B & $4.49 \pm 0.02$ & $0.41 \pm 0.05$ & $14.0 \pm 0.1$ & {}\\
P021400201103 & 58522.943 & 58522.955 & B & $4.46 \pm 0.04$ & $0.38 \pm 0.07$ & $14.6 \pm 0.2$ & {}\\
P021400201201 & 58523.314 & 58523.449 & B & $4.61 \pm 0.09$ & $0.60 \pm 0.14$ & $13.4 \pm 0.2$ & EP 1\\
P021400201501 & 58523.801 & 58523.948 & B & $4.54 \pm 0.02$ & $0.41 \pm 0.05$ & $14.3 \pm 0.1$ & {}\\
P021400201502 & 58524.072 & 58524.114 & B & $4.69 \pm 0.06$ & $0.43 \pm 0.09$ & $13.0 \pm 0.2$ & {}\\
P021400201503 & 58524.136 & 58524.246 & B & $4.57 \pm 0.07$ & $0.36 \pm 0.17$ & $9.7 \pm 0.6$ & {}\\
\hline
P021400201504 & 58524.275 & 58524.345 & B & $4.55 \pm 0.04$ & $0.45 \pm 0.13$ & $13.2 \pm 0.3$& {}\\
P021400201504 & 58524.372 & 58524.393 & C & $9.23 \pm 0.32$ & $7.21 \pm 0.93$ & $15.9 \pm 2.1$& EP 2\\
P021400201504 & 58524.393 & 58524.396 & B & $4.81 \pm 0.15$ & $1.24 \pm 0.32$ & $12.3 \pm 3.1$ & {}\\
P021400201505 & 58524.407 & 58524.743 & B & $4.56 \pm 0.02$ & $0.59 \pm 0.05$ & $13.5 \pm 0.1$ {}\\
\hline
P021400201506 & 58524.779 & 58524.799 & NO-Q & \nodata & \nodata & \nodata & {}\\
P021400201506 & 58524.799 & 58524.805 & B & $4.47 \pm 0.13$ & $0.79 \pm 0.32$ & $9.5 \pm 2.3$ & {}\\
P021400201506 & 58524.805 & 58524.808 & NO-Q & \nodata & \nodata & \nodata & {}\\
P021400201506 & 58524.848 & 58524.856 & B & $4.33 \pm 0.07$ & $0.43 \pm 0.12$ & $9.1 \pm 1.1$ & {}\\
P021400201506 & 58524.856 & 58524.861 & NO-Q & \nodata & \nodata & \nodata & {}\\
P021400201506 & 58524.861 & 58524.865 & B & $4.46 \pm 0.05$ & $0.43 \pm 0.08$ & $13.1 \pm 1.2$ & EP 3 \\
P021400201506 & 58524.865 & 58524.875 & NO-Q & \nodata & \nodata & \nodata & {}\\
P021400201507 & 58524.917 & 58524.922 & NO-Q & \nodata & \nodata & \nodata & {}\\
P021400201507 & 58524.922 & 58524.931 & B & $4.35 \pm 0.04$ & $0.35 \pm 0.11$ & $10.3 \pm 0.4$ &{}\\
P021400201507 & 58524.931 & 58524.933 & NO-Q & \nodata & \nodata & \nodata & {}\\
P021400201507 & 58524.933 & 58524.941 & B & $4.45 \pm 0.05$ & $0.44 \pm 0.09$ & $13.5 \pm 0.7$ & {}\\
\hline
P021400201508 & 58525.064 & 58525.176 & NO-Q & \nodata & \nodata & \nodata & {}\\
P021400201509 & 58525.198 & 58525.239 & NO-Q & \nodata & \nodata & \nodata & {}\\
P021400201510 & 58525.386 & 58525.468 & NO-Q & \nodata & \nodata & \nodata & {}\\
P021400201511 & 58525.500 & 58525.601 & NO-Q & \nodata & \nodata & \nodata & EP 4\\
P021400201512 & 58525.632 & 58525.737 & B & $4.35 \pm 0.04$ & $0.57 \pm 0.09$ & $13.0 \pm 0.1$ & {}\\
P021400201513 & 58525.787 & 58525.869 & B & $4.33 \pm 0.03$ & $0.49 \pm 0.07$ & $13.3 \pm 0.1$ & {}\\
\hline
P021400201601 & 58526.195 & 58526.232 & NO-Q & \nodata & \nodata & \nodata & {}\\
P021400201602 & 58526.379 & 58526.455 & NO-Q & \nodata & \nodata & \nodata & {}\\
P021400201603 & 58526.495 & 58526.595 & NO-Q & \nodata & \nodata & \nodata & {}\\
P021400201604 & 58526.625 & 58526.730 & NO-Q & \nodata & \nodata & \nodata & {}\\
P021400201605 & 58526.781 & 58526.815 & NO-Q & \nodata & \nodata & \nodata & EP 5\\
P021400201605 & 58526.815 & 58526.863 & B & $4.45 \pm 0.06$ & $0.44 \pm 0.07$ & $13.7 \pm 0.2$ &{}\\
P021400201606 & 58526.913 & 58526.995 & B & $4.57 \pm 0.04$ & $0.69 \pm 0.09$ & $14.6 \pm 0.1$ & {}\\
P021400201607 & 58527.022 & 58527.096 & B & $4.49 \pm 0.03$ & $0.47 \pm 0.08$ & $13.6 \pm 0.1$ & {}\\
\hline
P021400201701 & 58527.372 & 58527.448 & B & $4.56 \pm 0.04$ & $0.51 \pm 0.08$ & $13.2 \pm 0.1$ & {}\\
P021400201702 & 58527.487 & 58527.525 & C & $7.42 \pm 0.07$ & $4.13 \pm 0.32$ & $18.6 \pm 3.1$ & {}\\
P021400201702 & 58527.551 & 58527.588 & C & $10.21 \pm 0.16$ & $6.23 \pm 0.51$ & $19.2 \pm 3.4$ & EP 6\\
P021400201703 & 58527.618 & 58527.723 & B & $4.44 \pm 0.02$ & $0.41 \pm 0.03$ & $14.1 \pm 0.2$ & {}\\
P021400201704 & 58527.774 & 58527.856 & B & $4.34 \pm 0.03$ & $0.52 \pm 0.07$ & $14.3 \pm 0.1$ & {}\\
\hline
P021400202501 & 58539.134 & 58539.161 & B & $3.98 \pm 0.08$ & $0.45 \pm 0.15$ & $11.9 \pm 0.2$ & {}\\
P021400202502 & 58539.171 & 58539.534 & B & $4.16 \pm 0.02$ & $0.47 \pm 0.04$ & $12.6 \pm 0.1$ & {}\\
P021400202503 & 58539.559 & 58539.663 & B & $4.15 \pm 0.04$ & $0.55 \pm 0.08$ & $13.1 \pm 0.1$ & {}\\
P021400202504 & 58539.741 & 58539.795 &  B & $4.33 \pm 0.31$ & $0.71 \pm 0.86$ & $11.2 \pm 0.3$ & {}\\
P021400202504 & 58539.795 & 58539.804 & C & $7.78 \pm 0.40$ & $9.34 \pm 1.03$ & $20.1 \pm 6.8$ & EP 7\\
P021400202505 & 58539.828 & 58539.955 & C & $7.43 \pm 0.36$ & $11.77 \pm 0.88$ & $22.1 \pm 6.3$ & {}\\
P021400202505 & 58539.955 & 58539.940 & B & $4.08 \pm 0.02$ & $0.52 \pm 0.05$ & $12.7 \pm 0.4$ & {}\\
P021400202506 & 58539.993 & 58540.004 & B & $4.13 \pm 0.08$ & $0.39 \pm 0.18$ & $10.4 \pm 0.2$ & {}\\
P021400202507 & 58540.125 & 58540.463 & B & $3.96 \pm 0.03$ & $0.56 \pm 0.09$ & $11.3 \pm 0.1$ & {}\\
\enddata
\tablecomments{$^{a}$ The label `C' means type-C QPO are detected. Similarly, `B' corresponds to type-B QPO, `NO-Q' corresponds to NO-QPO.\\
$^{b}$ Centroid frequency of the Lorentzian component.\\
$^{c}$ Width of the Lorentzian component.\\
$^{d}$ Fractional rms of the LFQPO.\\
$^{e}$ The divided epochs.}
\end{deluxetable*}

\begin{deluxetable*}{cccccc}[!htbp]
\tablecaption{Best-fit spectral parameters of epochs with B-QPO and NO-QPO using model 1 {\tt\string tbabs}$\times$({\tt\string diskbb+gaussian+nthcomp+xillverCp)}
\label{B-nthcomp}}
\tablewidth{500pt}
\tabletypesize{\scriptsize}
\tablehead{
\colhead{Component} & \colhead{Parametes} & 
\multicolumn{2}{c}{EP 4} & \multicolumn{2}{c}{EP 5}\\
\cmidrule(rr){3-4} \cmidrule(rr){5-6}
\colhead{} & {} & \colhead{NO-Q} & \colhead{B-Q} & 
\colhead{NO-Q} & \colhead{B-Q}
} 
\startdata
TBabs & $N_{\rm H}(\times10^{22}\rm cm^{-2}$) &$0.913\pm0.009$ &$0.901\pm0.003$ &$0.908\pm0.006$ &$0.893_{-0.007}^{+0.009}$ \\
\hline
Diskbb & $kT_{\rm in} ({\rm keV})$ &$0.758\pm0.003$ &$0.752_{-0.005}^{+0.004}$ &$0.743\pm0.003$ &$0.734_{-0.007}^{+0.006}$ \\
{  }& norm &$27555_{-571}^{+621}$ &$27987_{-672}^{+421}$ &$30067_{-525}^{+582}$ &$28496_{-1007}^{+1203}$ \\
\hline
Gaussian & $E_{\rm line}({\rm keV})$ &$5.0_{-0.3}^{+0.2}$ &$4.3_{-0.4}^{+0.3}$ &$4.8_{-0.3}^{+0.2}$ &$3.7_{-0.6}^{+0.4}$ \\
{  } & $\sigma ({\rm keV})$ &$1.47_{-0.17}^{+0.16}$ &$1.87_{-0.15}^{+0.13}$ &$1.66_{-0.12}^{+0.13}$ &$2.03_{-0.19}^{+0.21}$ \\
{  } & norm &$0.27_{-0.07}^{+0.09}$ &$0.62_{-0.13}^{+0.16}$ &$0.38_{-0.07}^{+0.08}$ &$1.06_{-0.28}^{+0.31}$ \\
\hline
Nthcomp & $\Gamma$ &$2.26\pm0.02$ &$2.22_{-0.02}^{+0.03}$ &$2.21_{-0.02}^{+0.01}$ &$2.23_{-0.02}^{+0.03}$ \\
{  } & $kT_{\rm e} (\rm keV)$ &$105.1_{-34.9}^{+59.9}$ &$61.3_{-8.8}^{+20.1}$ &$102.9_{-27.5}^{+94.6}$ &$54.7_{-8.4}^{+20.1}$ \\
{  } & norm &$3.47_{-0.26}^{+0.19}$ &$3.86_{-0.13}^{+0.15}$ &$2.64_{-0.10}^{+0.11}$ &$4.80_{-0.14}^{+0.19}$ \\
\hline
Xillver & $A_{\rm Fe}$ &$2.40_{-1.02}^{+1.47}$ &$10.00_{-6.28}^{*}$ &$10.00_{-6.29}^{*}$ &$10.00_{-6.15}^{*}$ \\
{  } & log($\xi$) &$3.85_{-0.37}^{+0.29}$ &$3.47_{-0.37}^{+0.43}$ &$3.61_{-0.57}^{+0.44}$ &$3.84_{-0.29}^{+0.28}$ \\
{  } & norm &$0.017_{-0.008}^{+0.010}$ &$0.025_{-0.001}^{+0.003}$ &$0.002_{-0.001}^{+0.003}$ &$0.007_{-0.002}^{+0.008}$ \\
{} & probability & 3.24e-07 & 7.02e-08 & 0.0065 & 2.351e-07\\
\hline
{  } & $\chi^2/d.o.f$ &730.3/728 & 726.3/746 & 670.0/733 & 590.6/675  \\
\hline
Flux & Total &$16.68$ & $16.66$ & $15.50$ &$16.59$ \\
{ } & Diskbb &$12.78 $ & $12.36 $ & $12.51 $ &$11.35 $ \\
{  }& Gaussian &$0.29 $ & $0.43 $ & $0.31 $ &$0.64 $ \\
{ } & Nthcomp &$3.61$ & $3.87$ & $2.68$ &$4.60$ \\
\enddata
\tablecomments{Error bars are of 90\% confidence limit.\\
$^{*}$ means that the parameter pegs at its limit.\\
Flux unit is $10^{-8} \rm ergs/\rm cm^{2}/\rm s$ in the energy band 1–100 keV.\\
}
\end{deluxetable*}

\begin{deluxetable*}{cccccccc}
\tablecaption{Best-fit spectral parameters of epochs with B-QPO and C-QPO using model 1 {\tt\string tbabs}$\times$({\tt\string diskbb+gaussian+nthcomp+xillverCp)}
\label{C-nthcomp}}
\tablewidth{500pt}
\tabletypesize{\scriptsize}
\tablehead{
\colhead{Component} & \colhead{Parametes} & 
\multicolumn{3}{c}{EP 6} & \multicolumn{3}{c}{EP 7}\\ 
\cmidrule(rr){3-5} \cmidrule(rr){6-8}
\colhead{} & {} & 
\colhead{B-Q} & \colhead{C-Q} & 
\colhead{B-Q} & \colhead{B-Q} & 
\colhead{C-Q} &\colhead{B-Q}  
} 
\startdata
TBabs & $N_{\rm H}(\times10^{22}\rm cm^{-2}$) &$0.899_{-0.017}^{+0.018}$ &$0.870\pm0.011$ &$0.894\pm0.007$ &$0.859\pm0.006$ &$0.856_{-0.011}^{+0.004}$ &$0.867_{-0.007}^{+0.008}$\\
\hline
Diskbb & $kT_{\rm in} ({\rm keV})$ &$0.759_{-0.046}^{+0.010}$ &$0.683_{-0.007}^{+0.006}$ &$0.737_{-0.006}^{+0.005}$ &$0.689\pm0.004$ &$0.646_{-0.004}^{+0.007}$ &$0.685\pm0.004$ \\
{  }& norm &$23662_{-1425}^{+4510}$ &$32228_{-1289}^{+1376}$&$28206_{-797}^{+923}$ &$28437_{-617}^{+676}$ &$31670_{-734}^{+747}$ &$28920_{-681}^{+757}$ \\
\hline
Gaussian & $E_{\rm line}({\rm keV})$ &$4.6_{-3.8}^{+0.4}$ &$3.5_{-0.4}^{+0.4}$ &$3.9_{-0.5}^{+0.4}$ &$3.5_{-0.4}^{+0.3}$ &$3.4_{-0.3}^{+0.3}$  &$3.5_{-0.5}^{+0.3}$ \\
{  } & $\sigma ({\rm keV})$ &$1.13_{-0.36}^{+1.77}$ &$2.07_{-0.18}^{+0.18}$ &$1.96_{-0.16}^{+0.17}$ &$2.10_{-0.11}^{+0.13}$ &$2.24_{-0.10}^{+0.10}$ &$2.02_{-0.20}^{+0.16}$ \\
{  } & norm &$0.26_{-0.12}^{+1.56}$ &$1.16_{-0.27}^{+0.31}$ &$0.81_{-0.20}^{+0.28}$ &$0.87_{-0.14}^{+0.20}$ &$1.04_{-0.14}^{+0.16}$  &$0.69_{-0.19}^{+0.22}$ \\
\hline
Nthcomp & $\Gamma$ &$2.31_{-0.06}^{+0.02}$ &$2.15\pm0.02$ &$2.25_{-0.02}^{+0.03}$ &$2.20\pm0.01$ &$2.10\pm0.01$ &$2.22\pm0.02$ \\
{  } & $kT_{\rm e} (\rm keV)$ &$121.2_{-20.4}^{+23.1}$ &$84.3_{-18.1}^{+32.2}$ &$111.3_{-29.2}^{+108.1}$ &$108.5_{-23.9}^{+49.3}$ &$84.1_{-14.7}^{+22.8}$ &$142.2_{-48.8}^{+339.6}$ \\
{  } & norm &$4.86_{-4.86}^{+4.86}$ &$5.52_{-0.27}^{+0.30}$ &$4.19_{-0.16}^{+0.17}$ &$3.18_{-0.09}^{+0.09}$ &$2.90_{-0.11}^{+0.12}$  &$2.77_{-0.17}^{+0.13}$ \\
\hline
Xillver & $A_{\rm Fe}$ &$4.92_{-3.27}^{+4.88}$ &\nodata &$6.53_{-3.85}^{+-6.53}$ &$4.03_{-1.47}^{+2.81}$ &\nodata &$3.82_{-1.18}^{+1.95}$ \\
{  } & log($\xi$) &$4.15_{-1.24}^{+0.49}$ &\nodata &$3.50_{-0.29}^{+0.33}$ &$3.65_{-0.19}^{+0.19}$ &\nodata &$3.95_{-0.18}^{+0.30}$ \\
{  } & norm &$0.034_{-0.025}^{+0.012}$ &\nodata &$0.007_{-0.002}^{+0.006}$ &$0.006_{-0.002}^{+0.003}$ &\nodata &$0.011_{-0.004}^{+0.006}$ \\
{} & probability & 0.006882 & \nodata (0.75) & 2.25e-10 & 1.19e-21& \nodata & 4.29e-23\\   
\hline
{  } & $\chi^2/d.o.f$ &307.5/322 & 344/309 & 660.2/736 & 974.4/1017 & 682/615 & 792.0/919 \\  
\hline
Flux  & Total &$16.63$ &$15.35$ &$15.90$ &$12.04$ &$11.02$ &$11.60$ \\
{ }& Diskbb &$11.32 $ &$9.03 $ &$11.41 $ &$8.47 $ &$7.34 $ &$8.45 $ \\
{ }& Gaussian &$0.60 $ &$0.66 $ &$0.51 $ &$0.49 $ &$0.57 $ &$0.42 $ \\
{ } & Nthcomp  &$4.71$ &$5.67$ &$3.98$ &$3.08$ &$3.12$ &$2.73$\\
\enddata
\tablecomments{Error bars are of 90\% confidence limit.\\
$^{*}$means that the parameter pegs at its limit.\\
Flux unit is $10^{-8} \rm ergs/\rm cm^{2}/\rm s$ in the energy band 1–100 keV.
}
\end{deluxetable*}

\begin{deluxetable*}{lccccrcccc}[!htbp]
\tablecaption{Results of simultaneous spectral fits using model {\tt\string tbabs}$\times$({\tt\string diskbb+gaussian+nthcomp)}
\label{compare-nthcomp}}
\tablewidth{0pt}
\tabletypesize{\scriptsize}
\tablehead{
\colhead{} & \multicolumn{2}{c}{EP 4} & 
\multicolumn{2}{c}{EP 5} & \colhead{} &\multicolumn{2}{c}{EP 6} & \multicolumn{2}{c}{EP 7}\\ 
\cmidrule(rr){2-3} \cmidrule(rr){4-5} \cmidrule(rr){7-8} \cmidrule(rr){9-10}
\colhead{} & 
\colhead{NO-Q} & \colhead{B-Q} & 
\colhead{NO-Q} & \colhead{B-Q} & \colhead{} &
\colhead{B-Q} &\colhead{C-Q} & 
\colhead{B-Q} &\colhead{C-Q}
} 
\startdata
{Parameters} & \nodata & ${\chi^2_{red}/d.o.f}^{d}$  & \nodata &$\chi^2_{red}/d.o.f$ & {Parameters} & \nodata & $\chi^2_{red}/d.o.f$  & \nodata &$\chi^2_{red}/d.o.f$ \\
All tied &\nodata &	6326.3/1502	&\nodata &100662.1/1436 & All tied &\nodata & 32532.22/656 & \nodata & 81580.34/1657\\
${N_{\rm nthcomp}}^{a}$ & \nodata & 3712.9/1501 & \nodata & 13564/1435 &$N_{\rm diskbb}$ & \nodata & 7403.7/655 & \nodata & 16945.38/1656\\
${N_{\rm diskbb}}^{b}$ & \nodata & 1867.9/1500 & \nodata & 5265.4/1434 & $\Gamma $ &\nodata & 5312.17/654 & \nodata & 16403.51 /1655\\
${N_{\rm gaussian}}^{c}$ & \nodata & 1610.7/1499 & \nodata & 1658.7/1433 & $kT_{\rm in}$ &\nodata &1443.28/653 &	\nodata &2797.47/1654\\
{} & {} & {} & {} & {} & $N_{\rm nthcomp}$ &\nodata &763.98/652 & \nodata &2604.69/1653\\
\hline	
$N_{\rm nthcomp}$ &$3.46_{-0.10}^{+0.11}$ & $3.72\pm0.01$ & $2.56_{-0.08}^{+0.09}$ & $4.23\pm0.01$ &
$N_{\rm diskbb}$& $28089_{-1922}^{+3099}$ &$34772_{-347}^{+353}$ & $29730_{-815}^{+884}$& $34052_{-154}^{+155}$\\
$N_{\rm diskbb}$ & $27663_{-448}^{+485}$&$26967_{-26}^{+25}$&$29914_{-467}^{+516}$&$28036_{-31}^{+32}$ &
$\Gamma$ & $2.22_{-0.02}^{+0.02}$  & $2.15\pm0.01$ & $2.16\pm0.01$ &$2.108\pm0.003$\\
$N_{\rm gaussian} $ & $0.34_{-0.04}^{+0.05}$& $0.41\pm0.01$ & $0.40_{-0.05}^{+0.06}$ & $0.72\pm0.01$ & 
$kT_{\rm in}$ & $0.747_{-0.020}^{+0.013}$ & $0.684\pm0.002$ &$0.84\pm0.01$ &$0.684\pm0.002$\\
{} & {} & {} & {} & {}&	$N_{\rm nthcomp}$ & $1.22_{-0.09}^{+0.11}$ &$1.51\pm0.02$ & $0.87\pm0.02$&$0.931_{-0.007}^{+0.008}$
\enddata
\tablecomments{Error bars are of 90\% confidence limit.\\
$^{*}$ means that the parameter pegs at its limit\\
$^{abc}$ Normalization of \emph{nthcomp}, \emph{diskbb} and \emph{gaussian}, respectively.\\
$^{d}$ The chi-squared obtained after freeing the parameters.\\
\emph{Last row} presents results obtained after freeing the above parameters.
}
\end{deluxetable*}

\begin{deluxetable*}{llcccc}[!htbp]
\tablecaption{Best-fit spectral parameters of epochs with B-QPO and NO-QPO using model 2 {\tt\string tbabs$\times$({\tt\string diskbb+relxilllp)}}
\label{B-relxilllp}}
\tablewidth{500pt}
\tabletypesize{\scriptsize}
\tablehead{
\colhead{Component} & \colhead{Parametes} & 
\multicolumn{2}{c}{EP 4} & \multicolumn{2}{c}{EP 5}\\
\cmidrule(rr){3-4} \cmidrule(rr){5-6}
\colhead{} & {} & \colhead{NO-Q} & \colhead{B-Q} & 
\colhead{NO-Q} & \colhead{B-Q}
} 
\startdata
TBabs & $N_{\rm H}(\times10^{22}\rm cm^{-2}$) &$1.04_{-0.01}^{+0.02}$ &$1.02\pm0.01$ &$1.007\pm0.001$ &$1.03\pm0.01$ \\
\hline
Diskbb & $kT_{\rm in} ({\rm keV})$ &$0.762_{-0.002}^{+0.001}$ &$0.763_{-0.002}^{+0.003}$ &$0.746\pm0.001$ &$0.759\pm0.001$ \\
{  }& norm &$25078_{-142}^{+234}$ &$23760_{-413}^{+212}$ &$27333_{-58}^{+85}$ &$21770_{-178}^{+166}$ \\
\hline
Relxilllp & $h (GM/C^2)$ &$2.95_{-0.12}^{+0.08}$ &$3.18_{-0.12}^{+0.09}$ &$2.86_{-0.20}^{+0.04}$ &$3.21_{-0.07}^{+0.04}$ \\
{ } & $R_{\rm in}$ (ISCO) &$1.00_{*}^{+0.16}$ &$1.00_{*}^{+0.13}$ &$1.00_{*}^{+0.09}$ &$1.00_{*}^{+0.07}$ \\
 { } & $\Gamma$ &$2.40_{-0.02}^{+0.03}$ &$2.35\pm0.01$ &$2.39\pm0.01$ &$2.36\pm0.01$ \\
{ } & $a^*$ ($c{\rm J}/GM^2$) & \multicolumn{4}{c}{0.80 (fixed)}\\
{ } & $\theta$ (deg) & \multicolumn{4}{c}{36.5 (fixed)} \\
{  } & log($\xi$) &$3.77_{-0.06}^{+0.21}$ &$4.25_{-0.11}^{+0.07}$ &$4.23_{-0.04}^{+0.01}$ &$4.36_{-0.01}^{+0.01}$ \\
{  } & $A_{\rm Fe}$ &$6.03_{-1.24}^{+0.94}$ &$10.00_{-1.76}^{*}$ &$9.67_{-0.44}^{+0.03}$ &$10.00_{-0.71}^{*}$ \\
{  } & $E_{\rm cut ({\rm keV}})$ &$241.1_{-30.2}^{+61.8}$ &$219.1_{-24.5}^{+14.8}$ &$500.0_{-121.9}^{*}$ &$204.3_{-14.4}^{+12.3}$ \\
{  } & $R_{\rm ref}$ &$0.62\pm0.03$ &$0.63\pm0.03$ &$0.74_{-0.03}^{+0.01}$ &$0.67_{-0.02}^{+0.02}$ \\
{  } & norm &$1.43\pm0.15$ &$0.98_{-0.09}^{+0.08}$ &$1.06_{-0.04}^{+0.06}$ &$1.40_{-0.01}^{+0.12}$ \\
\hline
{  } & $\chi^2_{red}/d.o.f$ &786.5/729 & 755.2/747 & 724.6/734 & 617.1/676 \\
\hline
Flux  & Total  &$17.49$ &$17.46$ &$16.14$ &$17.53$ \\
{ } & Diskbb &$11.77 $ &$11.22 $ &$11.59 $ &$9.97 $ \\
{ }& Relxilllp & $1.41 $ &$1.72 $ &$1.42 $ & $2.22 $ \\
{ } & Cutoffpl &$4.32$ &$4.48$ &$3.18$ &$5.32$\\
\enddata
\tablecomments{Error bars are of 90\% confidence limit.\\
$^{*}$ means that the parameter pegs at its limit.\\
Flux unit is $10^{-8} \rm ergs/\rm cm^{2}/\rm s$ in the energy band 1–-100 keV.\\
}
\end{deluxetable*}

\begin{deluxetable*}{llcccccc}[!htbp]
\tablecaption{Best-fit spectral parameters of epochs with B-QPO and C-QPO using model 2 {\tt\string tbabs}$\times$({\tt\string diskbb+relxilllp)}
\label{C-relxilllp}}
\tablewidth{500pt}
\tabletypesize{\scriptsize}
\tablehead{
\colhead{Component} & \colhead{Parametes} & 
\multicolumn{3}{c}{EP 6} & \multicolumn{3}{c}{EP 7}\\ 
\cmidrule(rr){3-5} \cmidrule(rr){6-8}
\colhead{} & {} & 
\colhead{B-Q} & \colhead{C-Q} & 
\colhead{B-Q} & \colhead{B-Q} & 
\colhead{C-Q} &\colhead{B-Q}  
} 
\startdata
TBabs & $N_{\rm H}(\times10^{22}\rm cm^{-2}$) &$1.022_{-0.004}^{+0.007}$ &$1.004_{-0.012}^{+0.009}$ &$1.026_{-0.007}^{+0.008}$ &$0.975_{-0.003}^{+0.001}$ &$0.938_{-0.006}^{+0.003}$ &$0.969_{-0.007}^{+0.003}$\\
\hline
Diskbb & $kT_{\rm in} ({\rm keV})$ &$0.777_{-0.003}^{+0.004}$ &$0.710_{-0.003}^{+0.004}$ &$0.756\pm0.002$ &$0.706\pm0.001$ &$0.669_{-0.004}^{+0.002}$ &$0.699_{-0.002}^{+0.001}$ \\
{  }& norm &$19467_{-400}^{+336}$ &$22968_{-560}^{+532}$ &$22533_{-323}^{+326}$ &$22565_{-158}^{+139}$ &$25110_{-188}^{+102}$ &$23894_{-210}^{+100}$ \\
\hline
Relxilllp & $h (GM/C^2)$ &$4.06_{-0.31}^{+0.23}$ &$2.99\pm0.08$ &$4.54_{-0.11}^{+0.04}$ &$2.61_{-0.04}^{+0.03}$ &$2.51_{-0.03}^{+0.04}$ &$2.68_{-0.05}^{+0.03}$ \\
{ } & Rin (ISCO) &$1.0_{*}^{+0.9}$ &$1.0_{*}^{+0.4}$ &$1.0_{*}^{+0.1}$ &$1.00_{*}^{+0.02}$ &$1.00_{*}^{+0.06}$ &$1.00_{*}^{+0.04}$ \\
 { } & $\Gamma$ &$2.37_{-0.02}^{+0.01}$ &$2.30_{-0.02}^{+0.01}$ &$2.40\pm0.02$ &$2.38_{-0.01}^{+0.02}$ &$2.30_{-0.01}^{+0.02}$ &$2.38\pm0.01$ \\
{ } & $a^*$ ($c{\rm J}/GM^2$) & \multicolumn{6}{c}{0.80 (fixed)}\\
{ } & $\theta$ (deg) & \multicolumn{6}{c}{36.5 (fixed)} \\
{  } & log($\xi$) &$4.28_{-0.13}^{+0.18}$ &$4.21_{-0.04}^{+0.03}$ &$4.28_{-0.11}^{+0.13}$ &$4.28_{-0.03}^{+0.02}$ &$4.27_{-0.04}^{+0.02}$ &$4.27_{-0.06}^{+0.07}$ \\
{  } & $A_{\rm Fe}$ &$6.83_{-1.65}^{+0.96}$ &$5.01_{-0.33}^{+0.02}$ &$8.12_{-0.75}^{+1.42}$ &$8.03_{-0.37}^{+0.29}$ &$10.00_{-0.28}^{*}$ &$7.52_{-0.72}^{+0.82}$ \\
{  } & $E_{\rm cut ({\rm keV}})$ &$350_{-60}^{+25}$ &$500_{-51}^{*}$ &$372_{-63}^{+93}$ &$500_{-60}^{*}$ &$500_{-11}^{*}$ &$500_{-62}^{*}$ \\
{  } & $R_{\rm ref}$ &$0.64_{-0.12}^{+0.08}$ &$0.73_{-0.01}^{+0.09}$ &$0.63_{-0.04}^{+0.03}$ &$0.88_{-0.06}^{+0.08}$ &$1.05_{-0.01}^{+0.04}$ &$0.96_{-0.06}^{+0.04}$ \\
{  } & norm &$0.70_{-0.09}^{+0.08}$ &$0.52_{-0.05}^{+0.01}$ &$1.40_{-0.12}^{+0.11}$ &$1.75_{-0.10}^{+0.12}$ &$1.13_{-0.02}^{+0.01}$ &$1.17_{-0.08}^{+0.07}$ \\
\hline
{  } & $\chi^2_{red}/d.o.f$ &310.7/323 & 335.1/307 & 711.2/737 & 1120.3/1018 & 737.9/613 & 867.5/920 \\
\hline
Flux  & Total &$17.51$ &$16.13$ &$16.77$ &$12.61$ &$11.47$ &$12.12$ \\
{} & Diskbb &$10.03 $ &$7.41 $ &$10.21 $ &$7.45 $ &$6.41 $ &$7.51 $ \\
{  }& Relxilllp &$2.25 $ &$2.57 $ &$1.81 $ &$1.96 $ &$2.09 $ &$1.88 $\\
{ } & Cutoffpl  &$5.24$ &$5.79$ &$4.78$ &$3.22$ &$2.96$ &$2.72$
\enddata
\tablecomments{Error bars are of 90\% confidence limit.\\
$^{*}$ means that the parameter pegs at its limit.\\
Flux unit is $10^{-8} \rm ergs/\rm cm^{2}/\rm s$ in the energy band 1-–100 keV.\\
}
\end{deluxetable*}

\begin{deluxetable*}{lccccrcccc}[!htbp]
\tablecaption{Results of simultaneous spectral fits using model 2 {\tt\string tbabs}$\times$({\tt\string diskbb+relxilllp)}
\label{compare-relxilllp}}
\tablewidth{0pt}
\tabletypesize{\scriptsize}
\tablehead{
\colhead{} & \multicolumn{2}{c}{EP 4} & 
\multicolumn{2}{c}{EP 5} & \colhead{} &\multicolumn{2}{c}{EP 6} & \multicolumn{2}{c}{EP 7}\\ 
\cmidrule(rr){2-3} \cmidrule(rr){4-5} \cmidrule(rr){7-8} \cmidrule(rr){9-10}
\colhead{} & 
\colhead{NO-Q} & \colhead{B-Q} & 
\colhead{NO-Q} & \colhead{B-Q} & \colhead{} &
\colhead{B-Q} &\colhead{C-Q} & 
\colhead{B-Q} &\colhead{C-Q}
} 
\startdata
Parameters & \nodata & ${\chi^2_{red}/d.o.f}^{c}$  &\nodata & $\chi^2_{red}/d.o.f$ & Parameters & \nodata & $\chi^2_{red}/d.o.f$  &\nodata & $\chi^2_{red}/d.o.f$ \\
All tied & \nodata & 8460.37/1502 & \nodata & 111803.8/1436 &
All tied &\nodata & 29227.48/656 & \nodata & 64119.46/1658\\
$\Gamma$ & \nodata & 4063.71/1501 & \nodata & 14354.44/1435 &
$kT_{\rm in} ({\rm keV})$ & \nodata & 3404.34/655 & \nodata & 4316.91/1657\\
${N_{\rm diskbb}}^{a}$ & \nodata & 1742.65/1500 & \nodata & 8201.09/1434 & 
$h (GM/C^2)$ &\nodata & 1235.55/654 & \nodata &4022.04/1656 \\
$h (GM/C^2)$ & \nodata & 1628.05/1499 & \nodata & 3552.20/1433 &
$N_{\rm diskbb}$ &\nodata & 983.54/653 & \nodata &2569.73/1655\\
$kT_{\rm in} ({\rm keV})$ & \nodata & 1571.34/1498 & \nodata & 2096.51/1432 & 
$\Gamma$ ;$E_{\rm cut}$ &\nodata &700.66/651 & \nodata &2102.18/1653\\
${N_{\rm relxilllp}}^{b}$ & \nodata & \nodata & \nodata & 1980.67/1431 &
$R_{\rm ref}$ &\nodata &655.98/650 & \nodata & 2084.41/1652\\
$E_{\rm cut}$;$R_{\rm ref}$ & \nodata & \nodata & \nodata & 1378.33/1429 & &\nodata & \nodata & \nodata & \nodata\\
\hline
$\Gamma$ & $2.40_{-0.02}^{+0.03}$ & $2.38_{-0.02}^{+0.01}$ & $2.39 \pm 0.01$ & $2.32\pm0.01$ & 
$kT_{\rm in}$ & $0.777_{-0.003}^{+0.004}$ & $0.707\pm0.002$ & $0.706\pm0.001$ & $0.661\pm0.001$\\
$N_{\rm diskbb}$ & $25078_{-142}^{+234}$ & $23474_{-123}^{+119}$ & $27333_{-58}^{+85}$ & $21383_{-221}^{+135}$ &
$h$ & $4.42_{-0.11}^{+0.33}$ &$3.77_{-0.15}^{+0.14}$ &$2.61_{-0.04}^{+0.03}$ &$2.47_{-0.01}^{+0.02}$\\
$h (GM/C^2)$ & $2.95_{-0.12}^{+0.08}$ & $3.33\pm0.01$ & $2.86_{-0.20}^{+0.04}$ & $3.51_{-0.02}^{+0.03}$ &
$N_{\rm diskbb}$ & $19467_{-400}^{+336}$ & $23232_{-398}^{+406}$ &$22565_{-158}^{+139}$  &$26296_{-144}^{+152}$\\
$kT_{\rm in} ({\rm keV})$ & $0.762_{-0.002}^{+0.001}$ & $0.765\pm0.001$ & $0.746\pm0.001$ & $0.763\pm0.001$ &
$\Gamma$ & $2.37_{-0.02}^{+0.01}$ & $2.33_{-0.01}^{+0.01}$ & $2.38\pm0.01$ &$2.34\pm0.01$\\
$N_{\rm relxilllp}$ & \nodata & \nodata	& $1.06_{-0.04}^{+0.06}$ & $1.59_{-0.18}^{+0.11}$ &
$E_{\rm cut}$ & $350_{-60}^{+25}$ & $500^{*}$ & $500^{*}$ &$500^{*}$\\
$E_{\rm cut}$ & \nodata & \nodata & $500^{*}$ &  $159_{-10}^{+11}$&$R_{\rm ref}$ & $0.64_{-0.12}^{+0.08}$ &$0.92_{-0.11}^{+0.14}$ & $0.88_{-0.06}^{+0.08}$  &$1.12_{-0.05}^{+0.06}$\\
$R_{\rm ref}$ & \nodata & \nodata & $0.74_{-0.03}^{+0.01}$ & $0.62_{-0.02}^{+0.04}$ & &\nodata & \nodata & \nodata & \nodata\\
\enddata
\tablecomments{Error bars are of 90\% confidence limit.\\
$^{*}$ means that the parameter pegs at its limit.\\
$^{ab}$ Normalization of \emph{diskbb} and \emph{relxilllp}.\\
$^{c}$ The chi-squared obtained after freeing the parameters.\\
\emph{Last row} presents results obtained after freeing the above parameters.
}
\end{deluxetable*}

\end{document}